\documentclass[sigconf]{acmart}

\acmConference[Anonymous Submission to ACM CCS 2017]{ACM Conference on Computer and Communications Security}{Due 19 May 2017}{Dallas, Texas}
\acmYear{2017}

\usepackage{url}
\usepackage{breakurl}
\usepackage{listings}
\usepackage{enumitem}
\usepackage{subcaption}
\usepackage{xspace}
\usepackage{xcolor}
\usepackage{multirow}
\usepackage{makecell}
\usepackage{booktabs}
\usepackage{float} 
\usepackage{longtable}
\graphicspath{{../}{./}}

\usepackage{soul}

\newcommand{\ignore}[1]{}

\usepackage{pifont}

\newcommand{\looprepair}{\textsc{LoopRepair}\xspace}
\newcommand{\patchagent}{\textsc{PatchAgent}\xspace}
\newcommand{\codex}{\textsc{Codex}\xspace}

\newcommand{\rover}{\textsc{CodeRover-S}\xspace}

\author{Hulin Wang}
\affiliation{\institution{Arizona State University}\city{Tempe}\state{AZ}\country{USA}}
\email{hwang551@asu.edu}

\author{Zion Leonahenahe Basque}
\affiliation{\institution{Arizona State University}\city{Tempe}\state{AZ}\country{USA}}
\email{zbasque@asu.edu}

\author{Jie Hu}
\affiliation{\institution{Arizona State University}\city{Tempe}\state{AZ}\country{USA}}
\email{jiehu12@asu.edu}

\author{Ati Priya Bajaj}
\affiliation{\institution{Arizona State University}\city{Tempe}\state{AZ}\country{USA}}
\email{atipriya@asu.edu}

\author{Yibo Liu}
\affiliation{\institution{Arizona State University}\city{Tempe}\state{AZ}\country{USA}}
\email{yiboliu@asu.edu}

\author{Samuel Zhu}
\affiliation{\institution{Arizona State University}\city{Tempe}\state{AZ}\country{USA}}
\email{sjzhu@asu.edu}

\author{Giorgi Kobakhia}
\affiliation{\institution{Arizona State University}\city{Tempe}\state{AZ}\country{USA}}
\email{gkobakhi@asu.edu}

\author{Nikhil Chapre}
\affiliation{\institution{Arizona State University}\city{Tempe}\state{AZ}\country{USA}}
\email{nchapre@asu.edu}

\author{Will Rosenberg}
\affiliation{\institution{Arizona State University}\city{Tempe}\state{AZ}\country{USA}}
\email{whrosenb@asu.edu}

\author{Siddharth Mishra}
\affiliation{\institution{Arizona State University}\city{Tempe}\state{AZ}\country{USA}}
\email{smish149@asu.edu}

\author{Aditya Maheshbhai Gabani}
\affiliation{\institution{Arizona State University}\city{Tempe}\state{AZ}\country{USA}}
\email{agabani@asu.edu}

\author{Moritz Schloegel}
\affiliation{\institution{CISPA Helmholtz Center for Information Security}\city{Saarbrücken}\country{Germany}}
\email{moritz.schloegel@cispa.de}

\author{Adam Doupé}
\affiliation{\institution{Arizona State University}\city{Tempe}\state{AZ}\country{USA}}
\email{doupe@asu.edu}

\author{Yan Shoshitaishvili}
\affiliation{\institution{Arizona State University}\city{Tempe}\state{AZ}\country{USA}}
\email{yans@asu.edu}

\author{Ruoyu Wang}
\affiliation{\institution{Arizona State University}\city{Tempe}\state{AZ}\country{USA}}
\email{fishw@asu.edu}

\author{Tiffany Bao}
\affiliation{\institution{Arizona State University}\city{Tempe}\state{AZ}\country{USA}}
\email{tbao@asu.edu}

\newcommand{\sys}{\mbox{\textsc{Kumushi}}\xspace}

\usepackage{listings}
\usepackage{xcolor}

\definecolor{codegray}{rgb}{0.5,0.5,0.5}
\definecolor{codegreen}{rgb}{0.0,0.5,0.0}
\definecolor{codeblue}{rgb}{0.0,0.0,0.7}
\definecolor{codepurple}{rgb}{0.58,0.0,0.82}
\definecolor{codebg}{rgb}{0.97,0.97,0.97}

\usepackage{listings}
\usepackage{xcolor}

\definecolor{codegray}{rgb}{0.5,0.5,0.5}
\definecolor{codegreen}{rgb}{0.0,0.5,0.0}
\definecolor{codeblue}{rgb}{0.0,0.0,0.7}
\definecolor{codepurple}{rgb}{0.58,0.0,0.82}
\definecolor{codebg}{rgb}{0.97,0.97,0.97}

\lstdefinestyle{cstyle}{
    language=C,
    backgroundcolor=\color{codebg},
    basicstyle=\ttfamily\footnotesize,
    keywordstyle=\color{codeblue}\bfseries,
    commentstyle=\color{codegreen}\itshape,
    stringstyle=\color{codepurple},
    numberstyle=\tiny\color{codegray},
    numbers=left,
    numbersep=8pt,
    stepnumber=1,
    showstringspaces=false,
    breaklines=true,
    tabsize=4,
    frame=single,
    rulecolor=\color{codegray},
    captionpos=b,
    xleftmargin=2.5em,
    framexleftmargin=2em,
    morekeywords={U32,BYTE,MEM_read32,
                  mrb_bool,mrb_sym,uint16_t,
                  codegen_scope,FALSE,TRUE,
                  no_peephole,mrb_last_insn,
                  OP_JMPIF,OP_LOADI_1,OP_LOADINEG}
}
\begin{document}
\title{Root-Cause-Driven Automated Vulnerability Repair}

\begin{abstract}
	Recent LLM-based systems have made automated vulnerability repair increasingly practical, but two challenges remain. First, without strong signals about where a bug originates, repair agents drift toward shallow edits that silence the observed failure while leaving the underlying defect unresolved. Second, finding the root cause for bugs is hard: even developers familiar with the codebase frequently produce fixes that address symptoms rather than the root cause, and LLM-based agents, operating with noisier context and less program understanding, are no exception.

We present \sys\footnote{Kumushi: “Kumu” is Hawaiian for “root,” or “teacher,” and “shi” is Chinese for “teacher”; together, the term denotes a teacher of root cause.}, a root-cause-driven patching agent that addresses both challenges by combining diversified dynamic fault localization with evidence-weighted ranking to focus the LLM on the code most relevant to the defect. To rigorously measure whether \sys produces genuinely better patches, we also introduce a two-tier patch quality metric that pairs automated oracle validation with structured expert assessment of patches.

Evaluated on 178 C/C++ vulnerabilities, \sys substantially outperforms prior specialized repair agents under automated evaluation while matching a frontier commercial coding agent. 
Expert assessment then reveals differences that oracles cannot: \sys produces more root-cause fixes and fewer superficial patches, and is preferred in the majority of decisive pairwise comparisons. Together, these results demonstrate that progress in automated vulnerability repair requires not only stronger patching systems, but also richer evaluation methods capable of distinguishing genuine fixes from oracle-passing ones.
\end{abstract}

\ccsdesc{Security and privacy~Use https://dl.acm.org/ccs.cfm to generate actual concepts section for your paper}

\keywords{automated program repair, patch generation, root cause} 


\maketitle

\section{Introduction}%
\label{sec:intro}


Automated vulnerability repair is becoming increasingly critical as the volume of disclosed vulnerabilities continues to grow. 
Modern fuzzing infrastructure, such as OSS-Fuzz~\cite{ossfuzz}, has found thousands of bugs accompanied by reproducible proofs of concept (PoCs). 
More recently, efforts such as Project Glasswing~\cite{glaswing} have uncovered thousands of more vulnerabilities in critical software and show no sign of waning. 
The resulting backlog far outpaces developers' triaging capacity, creating a strong demand for tools that can quickly generate correct patches for vulnerable source code ~\cite{nvd}.

Recent advances in large language models (LLMs) have given rise to a new class of LLM-based patching agents~\cite{Nong2025APPATCHAA, Kim2025LogsIP, Ye2025WellBI, Yu2025PATCHAGENTAP, zhang2026fixing} that address this problem through an iterative workflow: 
the agent first \emph{localizes} the fault, then \emph{synthesizes} a candidate patch, and finally \emph{validates} it against replay-based test oracles comprising the original PoC and the developer test suite. 
If validation fails, the agent feeds the error signal back to drive the next round of localization and patch generation.
The loop repeats until all oracles pass.
Prior work has substantially improved this iterative loop by enabling agents to leverage program analysis techniques, such as static analysis~\cite{Yu2025PATCHAGENTAP}, symbolic execution~\cite{Ye2025WellBI}, and dynamic analysis approaches~\cite{zhang2026fixing}.

Despite recent advances in agentic patching, we observe that existing patching agents fail to reliably address the root causes of vulnerabilities and generate incomplete or incorrect patches.
First, root cause localization is fundamentally limited by the underlying analysis techniques.
Static approaches, such as \patchagent~\cite{Yu2025PATCHAGENTAP}, cannot precisely resolve indirect calls, leading to incorrect localization in cases like the heap out-of-bounds vulnerability in mruby~\cite{mruby}.
In contrast, dynamic approaches based on symbolic execution (e.g., \looprepair~\cite{Ye2025WellBI}) suffer from poor scalability and fail to handle this case in practice.
Second, LLM-based patch generation is unstable due to its sensitivity to noisy context.
Even when the correct fix location is identified, agents frequently drift and produce incorrect edits~\cite{context, Zhang2025AgenticCE}.
For example, when fixing a heap buffer overflow in hunspell~\cite{hunspell}, \patchagent correctly identifies the fix location but ultimately patches a nearby incorrect location, distracted by irrelevant context.
Both limitations are critical and must be resolved for identifying and fixing root causes effectively.

In this paper, we present \sys, a root cause-driven patching agent that addresses the issues identified above. 
To balance the advantages and limitations of static and dynamic analyses, \sys employs \emph{diversified dynamic fault localization}: instead of relying on static analysis alone, it collects execution evidence from a family of crash-triggering inputs, exposing root cause-correlated functions that a single trace would not surface. 
To prevent LLM's context drift, \sys employs an \emph{evidence-weighted ranking} stage that consolidates the candidate pool into a ranked set of Functions of Interest, which concentrates the agent's attention on the most crash-relevant code before patch generation begins.


We evaluate the effectiveness of \sys and existing approaches in repairing the root causes of software vulnerabilities on real-world open-source projects.
During the development of evaluation, we observe that assessing root cause repair is inherently challenging due to the lack of a \emph{reliable ground truth}.
This creates a chicken-and-egg problem: identifying the root cause of a vulnerability is itself non-trivial, which makes it difficult to determine whether a patch correctly addresses it.
Automatic root cause identification could be an interesting solution, but it remains an open research problem. 
Prior work usually treats developer patches as ground truth; however, studies have shown that a significant fraction of developer patches fail to address the underlying defect.
Even developers deeply familiar with the codebase frequently produce fixes that target symptoms rather than root causes~\cite{Jiang2022EvocatioCB, Li2017ALE, Sleuth}.

An alternative is to rely on replay-based oracles~\cite{Hu2025SoKAV}, which re-execute the original PoC and developer test suites to verify that a patch suppresses the observed crash while preserving existing functionality.
However, these oracles, analogous to dynamic testing in Programming Language theory, are complete but not sound\footnote{In PL theory, a system is sound if a program that passes the system's checking will not exhibit certain forbidden behaviors at runtime.}: a patch may eliminate the observed failure and pass all tests while leaving the underlying defect reachable through alternative execution paths~\cite{Sleuth}, even when test suites are augmented with automated techniques such as fuzzing.
This coverage gap limits the ability of oracle-based evaluation to distinguish true root-cause fixes from superficial repairs.
Therefore, accurately assessing patch quality requires deeper analysis beyond what automated oracles alone can provide. However, such analysis typically requires security expertise and substantial manual effort, which does not scale to comprehensive, real-world evaluations.
Consequently, it remains difficult to design an evaluation methodology that balances accuracy and practicality for large-scale, real-world systems.

To effectively evaluate \sys, we introduce a two-tier root-cause repair quality metric that combines automated oracle validation (``tier 1'') with structured expert assessment (``tier 2'') of patch completeness and correctness.

This framework enables a more rigorous evaluation of automated vulnerability patching systems such as \sys and provides stronger evidence of its ability to generate genuine root-cause fixes compared to existing state-of-the-art patching tools. 
%

We evaluated \sys on 178 C/C++ vulnerabilities against three state-of-the-art works from academia and industry: 
\patchagent~\cite{Yu2025PATCHAGENTAP}, \looprepair~\cite{Ye2025WellBI}, and OpenAI's \codex~\cite{openai_codex}. 
Under tier-1 oracle-based evaluation, \sys reaches a plausible patch rate of 85\%, outperforming \patchagent (72\%) and \looprepair (2\%), confirming the effectiveness of our root cause-driven localization. 


However, we observe that \sys and \codex are statistically indistinguishable under all three replay-based oracles: replaying the PoC, executing the test suite, and replaying fuzzer-generated variant PoCs.
This is a direct manifestation of the oracle coverage gap identified above. 
Both tools produce a similar percentage of plausible patches, but automated oracles alone cannot tell us which tool produces better patches.

Applying tier-2 expert assessment provides the deeper analysis needed to distinguish the two: \sys produces more root-cause fixes and fewer superficial patches than \codex. On the 144 bugs where both tools produced a plausible patch, human experts conducting an blind assessment preferred \sys on 68 (47\%), \codex on 39 (27\%), and judged 37 (26\%) cases as equivalent; a sign test on the 107 decisive pairs gives $p = 0.0065$, confirming that \sys produces a statistically significantly greater number of better patches when bug difficulty is controlled.

\medskip
\noindent
\textit{\textbf{Contributions:}} 
We make the following contributions: 

\begin{itemize}[leftmargin=*]
    \item We identify two recurring limitations in state-of-the-art LLM-based patching tools: static-only fault localization consistently fails to pinpoint root causes, and unfiltered context causes agents to pivot toward crash-site suppression rather than root-cause repair. 
    We further identify that existing oracle-based evaluation cannot reliably differentiate genuine root-cause fixes from superficial ones.
    \item We propose a root cause-driven patching approach that addresses both limitations through diversified dynamic fault localization and evidence-weighted FOI ranking, and implement and open-source \sys as a prototype of this approach.
    \item We conduct a two-tier evaluation on 178 C/C++ vulnerabilities. 
    \sys outperforms \patchagent by 18\% in plausible patches (tier 1), and human experts prefer \sys patches over those of \codex by a 20\% margin (tier 2). Notably, this preference gap is invisible to automated oracles and emerges only under human judgment.
\end{itemize}

\section{Background and Motivation}%
\label{sec:motiv}

We briefly discuss relevant background and investigate patch quality of existing approaches.

\subsection{LLM-based Automated Program Repair}

Software vulnerability repair has long been studied under the umbrella of Automated Program Repair (APR).
Early approaches relied on templates~\cite{Gao2016BovInspectorAI,Li2022RegexScalpelRE,Son2013FixMU,Kim2013AutomaticPG}, search-based mutations~\cite{Goues2012GenProgAG,Gao2019CrashavoidingPR}, semantic-based approaches\cite{gao2021beyond,Huang2019UsingSP,Gao2015SafeMF,Shariffdeen2021ConcolicPR}, and learning-based models~\cite{Zhou2024OutOS,Chi2020SeqTransAV} to synthesize candidate patches.
Recent advances in LLMs have changed the landscape, enabling repair tools that reason over code at the semantic level.
LLM-based patch generation follows two general approaches.
Prompt-based techniques~\cite{Nong2025APPATCHAA,Kim2025LogsIP} query the model directly with engineered prompts that supply the bug report, surrounding code context, and optional chain-of-thought scaffolding, then treat the model's response as a candidate patch.
%
On the other hand, agent-based techniques~\cite{Yu2025PATCHAGENTAP,Ye2025WellBI, zhang2026fixing} embed the LLM in an iterative loop equipped with tools for code navigation, compilation, and test execution, allowing the agent to gather evidence, propose patches, and refine them across rounds.
Within the agent-based approach, state-of-the-art systems differ chiefly in how they localize the root cause:
\patchagent equips the LLM with code-navigation tools (e.g., viewcode) and delegates localization entirely to the model. Starting from the sanitizer report, the agent retrieves source context iteratively and reasons about the defect statically, with no dynamic signal beyond the crash stack.
\looprepair localizes through concolic execution. Replaying the PoC under KLEE, it tracks how crash-relevant input bytes propagate through the program, retains the program points that carry the full taint, and ranks them by their distance to the crash. Each candidate is paired with a crash-free constraint, producing a semantically grounded set of suspect locations.
\rover localizes through dynamic tracing. The PoC is executed on an instrumented binary that captures the full caller/callee graph of the buggy run. Functions that executed but are absent from the crash stack are resolved to source symbols and appended to the sanitizer report, giving the agent execution-grounded seeds beyond what the bare stack trace provides.
More recently, general-purpose coding agents~\cite{openai_codex, claude_code, open_hand} have demonstrated strong competence on software-engineering benchmarks~\cite{swebench}, vulnerability reproduction~\cite{wang2025cybergym}, PoC generation~\cite{lee2025secbench}, code review~\cite{code_review_benchmark} and can be applied to vulnerability repair tasks~\cite{lee2025secbench,Yu2026PatchVI} with minimal task-specific configuration efforts.


\subsection{Patch Quality Evaluation in Practice}

A high-quality patch should repair the faulty invariant at the source of a vulnerability rather than merely mitigate the observed crash. 
Verifying this property is itself a research problem, but two evaluation methods are common in practice: manual expert review and dynamic oracle validation.

In expert reviews, a reviewer reconstructs the root cause from the crash report and PoC behavior and judges whether the patch addresses the underlying defect rather than guarding the crash site. 
The judgment is usually reliable but labor-intensive, and it does not scale to address the size of modern vulnerability backlogs. 

In terms of automated validation, the field usually relies on \emph{replay-based oracles}, including PoC replay, developer test suite re-execution, and fuzzing-based variant exploration~\cite{Hu2025SoKAV}. PoC replay checks if the reported crash no longer reproduces after the patch has been applied~\cite{Nong2025APPATCHAA, Ye2025WellBI}. Similarly, developer test suites are re-executed to check if documented behavior is preserved~\cite{Yu2025PATCHAGENTAP, Pearce2021ExaminingZV}. Fuzzing campaigns seeded from the original PoC can be used to explore nearby inputs and provide a stronger guard against variants that re-trigger the defect~\cite{Jiang2022EvocatioCB, Sleuth}.

We observe that these oracles are complete but not sound. 
A patch may pass every replay-based check while leaving the underlying defect reachable through alternative execution paths~\cite{Sleuth}. 
Two patches that pass the same oracle suite can therefore differ substantially in whether they repair the root cause.

\subsection{Preliminary Study}%
\label{sec:preliminary}

To understand the limitations of state-of-the-art LLM-based patching tools in practice, we conduct a preliminary study.
We first run 10 independent trials of \patchagent for each bug and identify 24 bugs that it fails to patch in every run.
Using this subset of bugs, we examine whether other state-of-the-art tools, including \looprepair~\cite{Ye2025WellBI}, exhibit similar limitations. 


We observe that \patchagent fails to locate or examine the correct fix location for 9 out of 24 bugs, suggesting that static analysis alone is insufficient to pinpoint root causes.
We use the developer patch as the definition of the correct fix location for these cases. 
While tools such as \looprepair incorporate dynamic information, for instance, constraints derived from symbolic execution, they face significant scalability and generality challenges.
In our evaluation, the KLEE symbolic execution engine used by \looprepair failed to execute on all 24 targets, producing no patches.
From these observations, we believe our system needs to use lightweight dynamic analysis that avoids the overhead of formal symbolic execution while still grounding the repair process in runtime evidence.

Throughout our experiment, we also observe that the \patchagent's reasoning frequently surfaces the right location, but lengthy call chains, unrelated source files, and tangential snippets accumulate alongside it. 
Under this noise, the LLMs drift from the correct fix sites and edits elsewhere.
In our study, 9 out of 24 bugs follow this drift pattern: the LLM initially examines the correct line, but the surrounding context pulls it away.
In another 6, it stays on the correct line but repeatedly produces patches that either fail to compile or fail to mitigate the crash, exhausting its repair budget.
These 15 cases suggest that even when the correct location is identified, ineffective context filtering remains a critical barrier to successful repair.

In summary, these two observations motivate \sys's design:

\begin{enumerate}[leftmargin=*, label=(\arabic*)]
\item \textit{Static analysis alone cannot pinpoint the root cause.}
\item \textit{Unfiltered context drowns out the correct fix site.} 
\end{enumerate}

\begin{figure*}[htbp]
  \centering
  \includegraphics[width=\textwidth]{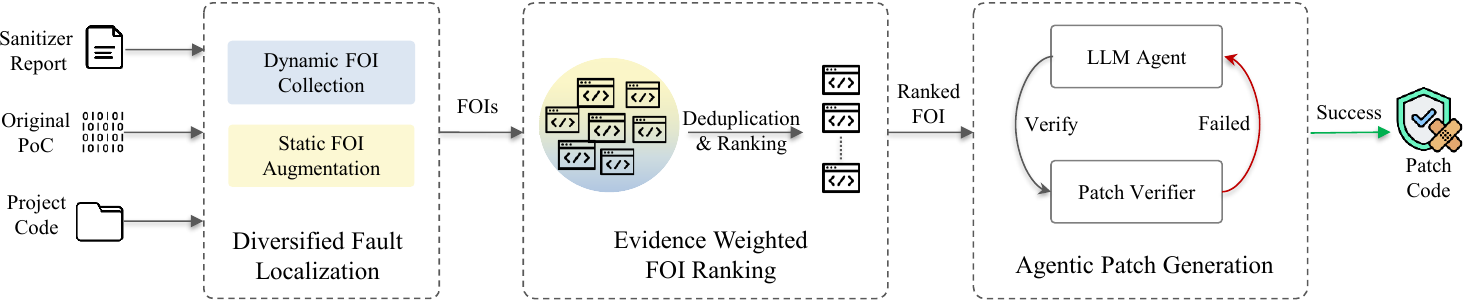}
  \caption{Overview of \sys. Given a sanitizer report, crash input, and project codebase, the Diversified Fault Localization stage produces a pool of Functions of Interest (FoIs); the Evidence-Weighted FoI Ranking stage consolidates and ranks them into a top-20 list; and the Agentic Patch Generation stage iteratively produces and verifies a patch, looping back with additional information on each failure until the patch is validated.}
  \label{fig:pipeline}
\end{figure*}

\newtheorem{workedexample}{Example}
\providecommand{\workedexampleautorefname}{Example}

\section{\sys: Root-Cause-Driven Patch Generation}%
\label{sec:design}

Motivated by these observations, we propose \sys, a dynamic-diversified, root-cause-focused, agentic patch generation framework that jointly addresses fault localization accuracy and patch quality.
As shown in Figure~\ref{fig:pipeline}, \sys accepts three inputs (a sanitizer report, the original proof-of-concept (PoC), and the project codebase) and operates through three tightly coupled stages.
First, \emph{Diversified Fault Localization} constructs a failure-relevant candidate pool.
It combines crash-stack extraction from the sanitizer report, dynamic function-call traces from fuzzer-generated crash variants of the original PoC, and static dependency analysis.
Second, \emph{Evidence-Weighted FoI Ranking} consolidates the pool into a focused top-20 set of Functions of Interest. 
It scores each candidate against crash-class-conditioned evidence tags, then applies a diversity-aware rerank so the patch agent receives distinct hypotheses across source files.
Finally, \emph{Agentic Patch Generation} consumes the ranked FoIs as context. 
The patch agent iteratively explores the codebase and synthesizes candidate patches, while an automated verifier checks each edit against oracles. The agent emits a patch only when all three checks pass.
The remainder of this section details each stage.

\subsection{Stage 1: Diversified Fault Localization}
\label{sec:design:fl}
Stage~1 collects a broad pool of functions likely correlated with the vulnerability's root cause. Stage~2 then consolidates, scores, and filters this pool into the focused top-20 FoI set that guides patch generation.
Motivated by the localization limitations identified in~\autoref{sec:motiv}, \sys addresses both the coverage gap of static-only approaches and the scalability limitations of heavyweight dynamic analysis through a two-step collection process.

\smallskip 
\noindent\textit{\textbf{Dynamic FoI Collection via Fuzzing.}} 
Rather than reasoning from a single crashing execution, \sys first diversifies the original PoC using AFL++~\cite{aflpp} (v4.40c) in crash exploration mode with a 12-hour time budget per bug. 
In crash exploration mode, the fuzzer takes the original crashing input as a seed and mutates it to generate new inputs, retaining only those that still crash the target. 
Aurora~\cite{aurora} observed that crash variants derived from the same seed tend to trigger the same underlying bug via different execution paths. The resulting family of variant inputs therefore exposes more root-cause-correlated functions than any single trace alone.
Crash exploration may occasionally produce inputs that trigger a different bug. Stage~2's evidence-weighted scoring down-weights functions that appear only in a minority of traces, mitigating such outliers.

For each variant in the diversified PoC family, \sys instruments the target binary with function-entry hooks and replays the variant under the instrumented binary. 
Every function executed on the path to the crash is emitted as a candidate FoI and enters the pool.
\sys additionally parses the ASan report of the original PoC and extracts every function named on the crash stack as a candidate FoI. Prior study found that the fix for roughly 60\% of cases patched a function on the crash stack~\cite{firefox_fix}.
The union of all per-variant execution traces and the crash-stack functions forms the dynamic candidate pool. 

\smallskip
\noindent\textit{\textbf{Static FoI Augmentation via Dependency Analysis.}}
Dynamic tracing captures functions that actually execute but may miss functions that influence the crash site through data dependencies not exercised by the available inputs. 
To augment FoI coverage, \sys applies CodeQL static dependency analysis~\cite{codeql}, but deliberately restricts it to the subset of functions that appear in the original PoC's ASan report. 
Applying static analysis to all dynamic candidates would re-introduce the over-approximation problem, exploding the candidate pool with functions only loosely related to the crash.
By anchoring static analysis to the ASan report subset, \sys obtains the precision benefits of crash-stack grounding while still recovering candidates that dynamic tracing alone would miss.

Concretely, CodeQL's intra-procedural dataflow query (\texttt{DataFlow\allowbreak::localFlow}) identifies variables whose values flow into the crash-line variables within the crash site function. 
A second CodeQL query enumerates all functions that read or write variables of matching name and access type. 
Whole-program dataflow is out of scope at our evaluation scale.

Beyond the local dataflow analysis, we take up to ten call-stack functions from the sanitizer report as \emph{anchors} and perform a bounded bidirectional breadth-first traversal of the call graph: upstream to callers and downstream to callees.
For spatial and temporal bugs we include allocation frames; for temporal bugs we also include free frames.
For each discovered function $f$, we track three signals: the number of anchors from which $f$ is reachable, $|\mathrm{anchors}(f)|$; the number of call edges landing on $f$, $\mathrm{edge}_{\mathrm{hits}}(f)$; and the shortest distance from any anchor to $f$, $\mathrm{min}_{\mathrm{depth}}(f)$. 
A function reached by many anchors, via many edges, at shallow depth is more likely to be implicated in the crash. 
We rank candidates lexicographically by $\langle -|\mathrm{anchors}|,\ -\mathrm{edge}_{\mathrm{hits}},\ \mathrm{min}_{\mathrm{depth}} \rangle$ and retain the top 300 to bound pool growth.
If too few candidates are returned, we widen the search by adding more call-stack, allocation-stack, and free-stack functions and merge the results.

After both steps, \sys merges the dynamically diversified candidates and the statically augmented candidates into a single pool of crash-correlated FoIs. Stage~2 then consolidates, scores, and ranks this pool.

\subsection{Stage 2: Evidence-Weighted FoI Ranking}
\label{sec:design:rank}

Stage~2 takes the merged candidate pool from Stage~1 and produces a focused, ranked top-20 FoI set that serves as the repair context for the patch agent. 
Each candidate in the pool carries one or more source tags indicating which collection step contributed it. 
The tags mark a function as appearing on the crash stack, in a per-variant dynamic execution trace, on an allocation or free stack from the sanitizer, at a sanitizer-reported object-origin site, or in the static variable-dependency results.
A candidate contributed by multiple steps carries multiple tags, and the scoring step below uses this overlap as its primary signal.

We first consolidate the pool, discarding functions that cannot plausibly be the defect site: fuzzing harnesses and entry wrappers (e.g., \texttt{LLVMFuzzerTestOneInput}), standard library primitives (\texttt{memcpy}, \texttt{malloc}, and similar), and candidates rooted in test or fuzzer directories.
Duplicates contributed by multiple steps are fused into a single record whose tag set is the union of the originals'.
We then score each surviving candidate. 
Tag weights depend on the crash class. 
For temporal errors such as use-after-free, the free stack carries the greatest weight; for contract violations such as null dereference, the crash stack carries it instead.
To prevent correlated sources from being counted multiple times, we first partition tags into evidence families.

For instance, free-stack, allocation-stack, and object-origin share one family because all three describe the same allocation event. Within a family \(f\), \sys aggregates the active tag weights using an OWA-style operator~\cite{owa}:

\begin{equation}
s_f = \min \Bigl(c_f,\; w_{(1)} + \alpha \sum_{i \geq 2} w_{(i)} \Bigr),
\end{equation}

where \(w_{(1)} \geq w_{(2)} \geq \dots\) are the bug-class weights of the active tags in that family, \(\alpha < 1\) discounts additional same-family tags to reflect correlation rather than independence, and \(c_f\) limits the maximum contribution of family \(f\). 

Intuitively, the strongest tag in a family determines the primary contribution, while the remaining tags provide only attenuated reinforcement, since tags from the same family often witness the same underlying fact. The cap \(c_f\) encodes the maximum confidence that the family can contribute in principle. 
For example, stack-trace evidence, which places the function directly on the crashing control flow, contributes more than call-trace evidence, which shows only that the function executed.

We then combine the family-level scores with a noisy-OR~\cite{pearl2014probabilistic}:

\begin{equation}
s = 1 - \prod_f (1 - s_f),
\end{equation}

\noindent under the standard assumption that evidence families provide independent support for defectiveness. 
\autoref{ex:rank_example} illustrates the resulting behavior. A single strong evidence family can substantially raise the final score, while weak or missing families do not pull down the contribution of stronger ones. The two operators serve complementary purposes. The OWA-style aggregation controls redundancy within a family, where multiple tags may be correlated observations of the same event. The noisy-OR combines support across families, where each family represents a distinct source by which a defect may manifest.

Given a score per candidate, the ranker produces the final list in two passes. We first rank by raw score, then apply a file-level diversity rerank to the tail. The rerank prefers candidates from source files not yet selected, so the patch agent receives distinct hypotheses rather than near-duplicates from the same file. We then truncate the list to match the patch agent's context budget.

\begin{workedexample}\label{ex:rank_example}
  Consider three use-after-free FoIs drawn from the same tag vocabulary
  (\emph{stack-trace} $0.85$, \emph{free-stack} $0.90$, \emph{alloc-stack} $0.65$,
  \emph{object-origin} $0.55$; $\alpha{=}0.25$, $c_{\text{crash}}{=}0.97$,
  $c_{\text{alloc}}{=}0.95$). By tag-family taxonomy, \emph{stack-trace} is a
  crash-stack tag, while \emph{free-stack}, \emph{alloc-stack}, and
  \emph{object-origin} are allocation-family tags. Each FoI fires its own subset.
  \begin{enumerate}[leftmargin=*,topsep=2pt,itemsep=1pt]
    \item \texttt{do\_close} fires \emph{stack-trace} (crash family) and both
      \emph{free-stack} and \emph{alloc-stack} (allocation family). It scores
      $\mathbf{0.9925}$:
      \begin{itemize}[leftmargin=*,topsep=2pt,itemsep=2pt]
        \item allocation family:
          $s_{\text{alloc}} = \min(0.95,\ 0.90 + 0.25\cdot 0.65) = 0.95$
        \item crash-stack family: $s_{\text{crash}} = \min(0.97,\ 0.85) = 0.85$
        \item noisy-OR: $s = 1-(1-0.85)(1-0.95) = 0.9925$
      \end{itemize}
    \item A FoI witnessed only by allocation-family tags --- with any combination
      of \emph{free-stack}, \emph{alloc-stack}, and \emph{object-origin} active
      --- cannot exceed $\mathbf{c_{\text{alloc}} = 0.95}$, since no crash-family
      term enters the noisy-OR.
    \item A FoI tagged only by \emph{stack-trace} on the crash stack scores
      $\mathbf{0.85}$.
  \end{enumerate}
  The ordering $0.85 < 0.95 < 0.9925$ makes the design choice visible: no amount
  of within-family evidence, even at saturation, overtakes a candidate witnessed
  across two independent families. \sys ranks breadth of evidence above depth.
  \end{workedexample}

\subsection{Stage 3: Agentic Patch Generation}
\label{sec:design:patch}

The patch agent is an LLM agent driven by a structured prompt and a small tool set. 
The prompt supplies the sanitizer report, the top-$20$ FoIs together with their source code, an explore-then-edit workflow, and fix-quality guidelines for patching vulnerabilities. 
We equip the agent with read-only tools to navigate the clang-indexed codebase, a string-substitution tool to edit source files, and a verifier to validate patches.
Every run operates on a freshly copied, git-baselined working directory so that the agent can explore the codebase and edit files without affecting the original codebase.

After each edit, the patch verifier applies three progressively stricter validation checks: compiling the patched project under AddressSanitizer, replaying the original crashing PoC, and re-executing the project's existing test suite. 
If any check fails, the verifier returns the failure output to the agent to guide its next exploration or edit.
The agent then can refine its hypothesis and try again.
Until the first time the patch passes all three checks, the run terminates and the agent emits the patch. 
The agent runtime caps each run at 150 turns to prevent infinite loops.



\section{Evaluation}%
\label{sec:evaluation}
We design experiments to study three research questions:
\begin{itemize}[leftmargin=*]
    \item \textbf{RQ1}: How do LLM-based patching agents compare under dynamic oracle-based evaluation? 
    \item \textbf{RQ2}: How do root-cause-driven and general-purpose LLM-based patching agents differ in patch comprehensiveness, and what repair strategies and failure modes characterize each? 
    \item \textbf{RQ3}: What factors drive patch quality differences between root-cause-driven and general-purpose LLM-based patching agents, and what do they reveal about the strengths and limitations of root-cause-anchored patching? 
\end{itemize}

\subsection{Experiment Setup}

\smallskip
\noindent\textit{\textbf{Tools.}}
We compare \sys against three state-of-the-art techniques that represent the spectrum of LLM-based patch generation.

\begin{itemize}[leftmargin=*]
    \item \textbf{\patchagent}~\cite{Yu2025PATCHAGENTAP} relies on the LLM alone for agentic fault localization and patch synthesis, iteratively refining candidate patches against the original PoC and developer test suite until the repair budget is exhausted.
    It uses no dynamic analysis beyond oracle replay validations.
    \item \textbf{\looprepair}~\cite{Ye2025WellBI} augments the LLM-based repair loop with symbolic execution: it runs KLEE on the original PoC's execution path to extract path constraints and taint traces that predict patch locations before delegating patch synthesis to an LLM.
    We configured it to use the same PoC-replay and test-suite oracle as PatchAgent for validation.
    \item \textbf{\codex}~\cite{openai_codex} is a general-purpose coding agent.
    We configure it with the original PoC, the developer test suite, and the project codebase, and prompt it to perform agentic patch generation, letting it autonomously decide how to explore the codebase and produce a patch.
    Unlike \sys, it receives no pre-computed fault-localization context and no fix-quality guidelines.
\end{itemize}

\noindent To ensure a fair comparison, all techniques use gpt-5.2-2025-12-11
~\cite{gpt}, the latest model while we ran the experiment, as the underlying large language model backend. 


\smallskip
\noindent\textit{\textbf{Dataset.}} We evaluate on the 178-bug benchmark introduced by Yu et al.~\cite{Yu2025PATCHAGENTAP}, which spans 30 open-source C/C++ projects ranging from 5k to 1M source lines of code (SLOC).
Each bug ships with an AddressSanitizer (ASan) crash report, a crash input, and a script to run test suites.
We group ASan crash labels into five classes based on the ASan report label; Table~\ref{tab:benchmark} gives the class definitions.
\autoref{app:bench} lists the full set of targets.

\begin{table}[H]
  \centering
  \small
  \setlength{\tabcolsep}{4pt}
  \caption{Overview of the 178-bug benchmark. 
}
  \label{tab:benchmark}
  \begin{tabular}{@{}l r r r r r r r@{}}
  \toprule
  \textbf{Size (SLOC)} & \textbf{\#Proj.} & \textbf{Spatial} & \textbf{UAF} & \textbf{NPD} & \textbf{Num} & \textbf{Other} \\
  \midrule
  Small ($<$50\,k)      & 10 &  25 &  6 &  2 & 1 & 5 \\
  Medium (50\,k–250\,k)      & 13 &  40 &  6 &  5 & 4 & 7 \\
  Large ($>$250\,k)     &  7 &  52 & 11 & 11 & 1 & 2 \\
  \midrule
  \textbf{Total}             & \textbf{30} & \textbf{117} & \textbf{23} & \textbf{18} & \textbf{6} & \textbf{14} \\
  \bottomrule
  \multicolumn{7}{l}{\footnotesize{}\textbf{Spatial}: heap-, stack-, or global-buffer overflow, or heap out-of-bound read.} \\
  \multicolumn{7}{l}{\footnotesize{}\textbf{UAF}: use-after-free, double-free, and invalid-free. } \\
  \multicolumn{7}{l}{\footnotesize{}\textbf{NPD}: null-pointer dereference.} \\
  \multicolumn{7}{l}{\footnotesize{}\textbf{Num}: divide-by-zero, integer overflow, and invalid bit-shift. } \\
  \multicolumn{7}{l}{\footnotesize{}\textbf{Other}: sanitizer-unclassified crashes including unknown-read/write, } \\
  \multicolumn{7}{l}{\footnotesize{}\phantom{\textbf{Other}: }SEGV-on-unknown-address, and API-contract violations.} \\
  \end{tabular}
\end{table}

\smallskip
\noindent\textit{\textbf{Prompt.}} The prompt for \sys's patching agent includes the root-cause analysis summary, the sanitizer report, the overall workflow, and the patching guidelines.
\autoref{appendix:llm-prompt} shows the complete prompt.

\smallskip
\noindent\textit{\textbf{Token usage.}}
\autoref{tab:repair-tokens} reports per-bug mean and median token consumption and estimated cost for each system.
All four systems use GPT-5.2 under their default repair budgets.

\begin{table}[H]
\centering
\small
\caption{Token usage and estimated cost (using GPT-5.2).}
\label{tab:repair-tokens}
\begin{tabular}{@{}lrrrr@{}}
  \toprule
   & \multicolumn{2}{c}{\textbf{Mean}} & \multicolumn{2}{c}{\textbf{Median}} \\
  \cmidrule(lr){2-3} \cmidrule(lr){4-5}
  \textbf{System} & \textbf{Tokens} & \textbf{Cost} & \textbf{Tokens} & \textbf{Cost} \\
  \midrule
   \patchagent &    44{,}899 & \$0.16 &  43{,}255 & \$0.16 \\
   \looprepair &    83{,}936 & \$0.30 &  50{,}046 & \$0.18 \\
   \codex      &   239{,}848 & \$0.86 & 130{,}830 & \$0.47 \\
   \sys        & 1{,}120{,}538 & \$4.02 & 517{,}992 & \$1.86 \\
  \bottomrule
\end{tabular}
\end{table}

\subsection{Root Cause Assessment Metric}%
\label{sec:metric}

To rigorously measure the performance of each tool and compare the generated patches comprehensively, we introduce a two-tier patch quality metric that combines the scalability of automated oracles with the precision of human assessment. 

Our two-tier metric design considers both evaluation precision and scalability.
Automated oracles such as PoC replay, test suites, and fuzzing are cheap to run at scale but passing them does not prove the underlying defect has been fixed~\cite{Hu2025SoKAV}, as a patch may suppress the observed failure while leaving the root cause unaddressed. 
Human review can determine whether a patch fixes the root cause, but applying it to every generated patch is prohibitively expensive, taking ~16 minutes per patch~\cite{Tao2014AutomaticallyGP}.
Our metric addresses this tension by running automated oracles first to eliminate failed patches, then forwarding only the surviving plausible patches to human raters who assess whether the root cause has been resolved.
Since the oracle stage requires only the patch, the crash report, and the program code base, our metric works for patches from any repair system.


\subsubsection{Tier 1: Dynamic Oracle Validation}%
\label{sec:tier1}

Three oracles cover the most common checks for validating LLM-generated patches~\cite{Hu2025SoKAV}:
\emph{PoC replay} confirms that a reported crash no longer triggers. 
The \emph{developer test suite} checks that documented behavior still holds. 
\emph{AFL++ crash exploration}, seeded from the PoC and run for 12 hours on the project's harness, detects patches that silence the reported crash but leave the defect reachable through mutated inputs.

We categorize each generated patch into one of three outcomes under
progressively stricter dynamic oracles:
\begin{itemize}[leftmargin=*]
    \item \textbf{No Patch}: the patch fails to compile, or mitigate the original PoC crash, or pass the developer's existing test suite.
    \item \textbf{Partial Patch}: the patch passes compilation, PoC replay, and the developer test suite, but does not suppress all variant crashes discovered through 12 hours of AFL++ crash-exploration fuzzing against the project's own harness.
    \item \textbf{Plausible Patch}: the patch passes every dynamic
    check above, including suppression of all AFL++-discovered variant crashes within the 12-hour exploration budget.
\end{itemize}

\subsubsection{Tier 2: Structured expert assessment}%
\label{sec:expert-eval}
We forward patches that pass all three dynamic oracles to human raters, who evaluate each patch along two orthogonal axes and then, for \sys and \codex pairs, apply a pairwise comparison rubric. The primary axis captures whether the patch addresses the underlying defect:

\begin{itemize}[leftmargin=*]
    \item \textbf{Root-Cause Fix}: the patch directly repairs the underlying defect identified from the crash report and PoC, eliminating the faulty invariant at its source. 
    \item \textbf{Symptom Fix}: the patch prevents or reduces the crash without addressing the underlying defect, leaving related execution paths potentially vulnerable. 
\end{itemize}

The secondary axis is a boolean flag, \texttt{has\_unrelated\_changes}, set whenever the diff contains modifications orthogonal to the root cause. 
We track this flag because LLM-based tools occasionally edit code beyond the scope of the root cause: reformatting, refactoring adjacent functions, or making drive-by changes. 
These edits inflate apparent productivity, add review burden, and risk regressions.
Surfacing such edits as a separate signal lets us distinguish a clean root-cause fix from a correct fix bundled with unwanted changes. 

We recruited eight raters (graduate students with research experience in software security and with more than 1 year of vulnerability research experience).
%
For each bug, three raters first reconstruct the runtime environment in Docker to run the PoC, then identify the root cause from the crash report and observed PoC behavior. 
Both \sys and \codex generated patches against identical Docker environments; only the patch-generation agent differed, controlling for confounders in the reproduction setup. 
We consult the original developer patch as a reference only and do not treat it as ground truth, since upstream fixes are themselves frequently incomplete or symptom-level~\cite{Jiang2022EvocatioCB,Sleuth}.

To mitigate source bias, we anonymize \codex and \sys outputs as \texttt{tool\_a} and \texttt{tool\_b}; the mapping is disclosed only after labeling is complete.
Labeling proceeds in independent rounds: each rater labels every patch in isolation without access to the others' judgments, and raters rotate across batches of bugs across rounds.

In each round, the rater assigns a primary label (root-cause fix or symptom fix) and writes a justification that references the identified root cause. 
After a dedicated scan for out-of-scope edits, the rater sets the \texttt{has\_unrelated\_changes} flag.
Raters record every flagged edit in a shared spreadsheet, along with their rationales.
Only after both patches have been labeled in isolation raters apply the \hyperref[rubric]{pairwise comparison rubric}.
Correctness takes priority: a root-cause fix outranks a symptom fix. 
When both patches repair the root cause, the rater breaks ties first by \emph{semantic precision}, preferring the patch that targets the faulty invariant without broadening scope. 
The number of unrelated changes serves as the second tiebreaker.
When neither patch reaches the root cause in our evaluation, we fall back to comprehensiveness, preferring the patch that covers more affected execution paths, edge cases, and related failure modes.
Raters label clear dominance cases directly, reserve the rubric for close calls, and log all rationales for post-hoc audit.
After three rounds, we use majority vote to determine the final label and provide inter-rater reliability statistics in \autoref{tab:fleiss-kappa}.

\phantomsection
\label{rubric}
\textit{Pairwise Comparison Rubric.}
The rubric is lexicographic on:
\begin{enumerate}[leftmargin=*]
\item \textbf{Correctness.} When one patch repairs the root cause and the other only suppresses the reported crash, prefer the root-cause fix. 
When both patches repair the root cause, prefer the one with stricter semantic alignment to the defect, which is the patch that targets the faulty invariant without broadening scope. 
Unrelated changes weigh against a patch under\texttt{has\_\allowbreak{}unrelated\_\allowbreak{}changes};
when counts are close, raters exercise judgment rather than apply a fixed threshold.
\item \textbf{Comprehensiveness.} When neither patch repairs the root cause, prefer the symptom fix that covers more of the relevant execution paths, edge cases, and related failure modes stemming from the same defect. 
Unrelated changes again weigh against the patch under the same judgment-based weighting.
\end{enumerate}

\subsection{RQ1: Patching Capability}

To answer RQ1, we run \sys against three baselines on 178 C/C++ vulnerabilities under three dynamic oracles: PoC replay, developer test suite, and 12h of AFL++ variant-crash exploration per bug.
Table~\ref{tab:patch_cap} reports the full outcome distribution.
We classify each patch as No Patch, Partial Patch, or Plausible Patch.
\sys substantially outperforms \patchagent and \looprepair, confirming the effectiveness of root-cause-driven localization.
\sys and \codex, however, are statistically indistinguishable under all three oracles, revealing a ceiling on what oracle-based metrics can discriminate at the frontier.



\begin{table}[H]
  \centering
  \caption{Patch quality across 178 C/C++ vulnerabilities, judged by dynamic oracles.}
  \label{tab:patch_cap}
  \small
  \begin{tabular}{lccc}
  \toprule
  \textbf{Tool} & \textbf{No Patch} & \textbf{Partial Patch} & \textbf{Plausible Patch} \\
  \midrule
  \patchagent  & 13.5\% (24)\phantom{h}  & 14.6\% (26)  & 71.9\% (128) \\
  \looprepair  & 97.2\% (173) & 0.6\%  (1)   & 2.2\%  (4)\phantom{h}   \\
  \codex       & 2.2\%  (4)\phantom{h}   & 11.8\% (21)  & 86.0\% (153) \\
  \sys         & 1.7\%  (3)\phantom{h}   & 12.9\% (23)  & 85.4\% (152) \\
  \bottomrule
  \end{tabular}
  \end{table}

\smallskip
\noindent\textit{\textbf{Per-tool breakdown.}} 
\patchagent produces no patch on 13.5\% of bugs and only partial patches on another 14.6\%, for a combined failure rate of 28.1\%.
The partial-patch cases suppress the reported crash but fail under AFL++ variants; the no-patch cases concentrate on bugs where static context retrieval returned code irrelevant to the defect.

\looprepair fails on 97.2\% of bugs.
Its symbolic-execution component (KLEE) compiled on 68 of the 178 targets and produced usable output on only 5, leaving \looprepair without localization input on the rest of the benchmark.
Beyond this implementation limitation, symbolic execution is more expensive than the dynamic tracing \sys uses.
Even on targets where KLEE succeeds, its per-target cost cannot scale to the diversified multi-input tracing \sys performs across the AFL++ variant family.


\begin{figure}[htbp]
    \centering
    \includegraphics[width=0.7\columnwidth]{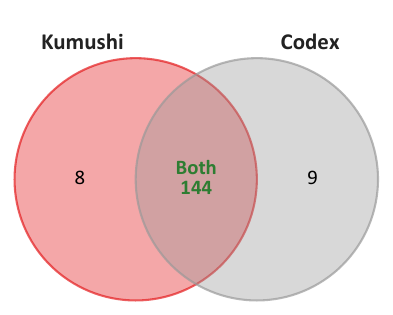}
    \caption{Comparison of \sys and \codex on plausible patch classification. The two systems overlap on 144 bugs; \sys uniquely patches 8 and \codex uniquely patches 9.} \label{fig:kumushi_codex_patch_classification}
\end{figure}

\sys and \codex produce nearly identical outcome distributions under all three oracles (\autoref{tab:patch_cap}).
Among the 161 bugs where either tool produced a plausible patch, only 144 are shared: \sys patches 8 bugs that \codex misses, and \codex patches 9 that \sys misses (\autoref{fig:kumushi_codex_patch_classification}).
The aggregate counts are indistinguishable, but the tools fix different sets of bugs.
The indistinguishability of \sys and \codex under all three oracles is not evidence of equivalent patch quality; it is evidence that oracle-based metrics cannot separate frontier tools at this level.
RQ2 applies tier-2 human assessment to examine what the oracles cannot see.

\subsection{RQ2: Patch Comprehensiveness}

To answer RQ2, we examine whether \sys and \codex differ in the quality of the plausible patches they produce, a dimension that the three oracle metrics in RQ1 cannot capture.
Three raters independently label all 305 plausible patches (152 from \sys and 153 from \codex) along two axes following the tier-2 protocol in Section~\ref{sec:expert-eval}: a root-cause vs.\ symptom judgment and a boolean \texttt{has\_unrelated\_changes} flag.
We then characterize the repair strategies each tool employs through open coding, yielding eight root-cause fix strategies and nine symptom-fix failure strategies. 

Tier-2 labeling reveals that \sys produces more root-cause fixes and fewer symptom fixes than \codex, and that the two tools diverge in the kind of repair they produce when they fall short of the root cause.




\smallskip
\noindent\textit{\textbf{Aggregated results.}}
Across 305 labeled patches, \sys produces more root-cause fixes than \codex (84.9\% vs.\ 77.8\%; 129 vs.\ 119, \autoref{tab:patch_qual}), indicating that \sys generates more comprehensive and correct patches than \codex overall.
The root-cause fix gap falls short of $p < 0.05$ at this sample size; the unpaired aggregate comparison has limited statistical power, and the per-bug paired analysis in RQ3 is better suited to detect the difference.


\sys also produces more patches containing unrelated edits than \codex (28.3\% vs.\ 21.6\%; 43 vs.\ 33, Table~\ref{tab:patch_qual}).
This higher rate does not signal lower patch quality from \sys.
Within each tool, the unrelated-edit rate is consistent across root-cause fixes and symptom fixes (29.5\% vs.\ 21.7\% for \sys; 22.7\% vs.\ 17.6\% for \codex), with neither within-tool difference reaching statistical significance ($p > 0.05$ for both).
Unrelated edits therefore reflect a stable property of each tool's editing behavior rather than a signal of patch incorrectness, and the higher unrelated-edit rate in \sys does not account for its root-cause fix advantage over \codex.

\begin{table}[H]
\centering
\small
\caption{Patch quality and unrelated change rate per tool.}
\label{tab:patch_qual}
\begin{tabular}{@{}llrrrr@{}}
\toprule
\textbf{Tool} & \textbf{Patch Quality} & 
\textbf{Count} & 
\textbf{\makecell[r]{With\\Unrelated}} & 
\textbf{\makecell[r]{Without\\Unrelated}} \\
\midrule
\multirow{3}{*}{\codex}
  & root-cause fix  & 119 & 27 &  92 \\
  & symptom fix     &  34 &  6 &  28 \\
  & \textbf{total}  & \textbf{153} & \textbf{33} & \textbf{120} \\
\midrule
\multirow{3}{*}{\sys}
  & root-cause fix  & 129 & 38 &  91 \\
  & symptom fix     &  23 &  5 &  18 \\
  & \textbf{total}  & \textbf{152} & \textbf{43} & \textbf{109} \\
\bottomrule
\end{tabular}
\end{table}



\smallskip
\noindent\textit{\textbf{Repair strategy analysis.}}
We next group each patch by its primary repair strategy through open coding, yielding eight root-cause fix strategies and nine symptom-fix strategies (full definitions in \autoref{app:repair_strategies}).
\autoref{tab:root-cause-taxonomy} and \autoref{tab:symptom-taxonomy} present the strategy name and per-tool counts for root-cause fixes and symptom fixes, respectively.

Among each tool's root-cause fixes (\autoref{tab:root-cause-taxonomy}), the per-tool category distributions are broadly similar: both lean on bound checks (35\% for \codex, 33\% for \sys) and API-contract enforcement.
The one notable difference is invariant validation, where \sys produces three times as many fixes as \codex (12 vs.~4).
This is consistent with \sys's diversified fault localization more often surfacing the upstream producer where the invalid input enters the system, rather than anchoring at the downstream consumer where the crash surfaces.

\begin{table}[htbp]
\centering
\small
\caption{Root-cause fix strategies ($n = 248$ after audit).}
\label{tab:root-cause-taxonomy}
\begin{tabular}{@{}lrrr@{}}
\toprule
\textbf{Category}      & \textbf{\codex} & \textbf{\sys} & \textbf{Total} \\
\midrule
Bound check            &  42 &  43 &  85 \\
API contract           &  21 &  21 &  42 \\
Size arithmetic        &  20 &  19 &  39 \\
Control-flow fix       &  13 &  12 &  25 \\
Ownership repair       &  11 &  12 &  23 \\
Invariant validation   &   4 &  12 &  16 \\
State sync             &   7 &   7 &  14 \\
Type representation    &   1 &   3 &   4 \\
\midrule
Total                  & 119 & 129 & 248 \\
\bottomrule
\end{tabular}
\end{table}

Among symptom fixes (\autoref{tab:symptom-taxonomy}), the two tools diverge more sharply.
\codex produces nearly twice as many crash-site guards as \sys (13 vs.\ 7) and is alone in producing trigger blocks (3 vs.\ 0); both are narrow patches that silence the observed failure without enforcing the general invariant.
\sys's symptom fixes spread more evenly across categories with no single failure mode dominating, suggesting that when \sys does not reach the root cause it does so for more varied reasons rather than consistently drifting to the crash site.

Together, the aggregate gap and the categorical asymmetries agree.
\sys more often repairs the defect at its source, while \codex more often patches at the crash site.
RQ3 confirms this difference with a paired within-bug comparison and illustrates its qualitative character through case studies.

\begin{table}[htbp]
\centering
\small
\caption{Symptom-fix strategies ($n = 57$ after audit).}
\label{tab:symptom-taxonomy}
\begin{tabular}{@{}lrrr@{}}
\toprule
\textbf{Category}         & \textbf{\codex} & \textbf{\sys} & \textbf{Total} \\
\midrule
Crash-site guard          & 13 &  7 & 20 \\
Downstream guard          &  4 &  5 &  9 \\
Value masking             &  4 &  4 &  8 \\
Corrupt-state tolerance   &  4 &  4 &  8 \\
Trigger block             &  3 &  0 &  3 \\
Incomplete coverage       &  2 &  1 &  3 \\
Lifetime masking          &  2 &  1 &  3 \\
Wrong location            &  1 &  1 &  2 \\
Ineffective guard         &  1 &  0 &  1 \\
\midrule
Total                     & 34 & 23 & 57 \\
\bottomrule
\end{tabular}
\end{table}

\smallskip
\noindent\textit{\textbf{Inter-rater reliability.}}
For each label we report three statistics in \autoref{tab:fleiss-kappa}: Fleiss's $\kappa$~\cite{fleiss1971measuring}, the raw observed agreement (mean of the three pairwise rater-agreement rates), and the prevalence of the positive class.
We compute the patch quality and unrelated-change labels over the 161-bug union and the pairwise winner label over the 144-bug intersection where both tools produced a plausible patch.
We treat abstentions as their own category.

On \emph{patch quality}, raw agreement is $0.81$/$0.86$ (\codex/\sys) at a positive-class (root-cause fix) prevalence of $0.75$/$0.83$, giving $\kappa = 0.57$/$0.62$.
The relatively low $\kappa$ reflects prevalence compression rather than rater disagreement~\cite{byrt1993bias}: when the positive class dominates, chance agreement is high and $\kappa$ deflates even when raters largely agree.
The root-cause vs.\ symptom-fix label, assigned after raters first identified each bug's root cause, is reproducible across rounds.
%
On \emph{unrelated change}, raw agreement is $0.84$/$0.74$ at a near-balanced prevalence of $0.25$/$0.32$, giving $\kappa = 0.64$/$0.47$.
The moderate $\kappa$ for \sys unrelated changes reflects the inherent subjectivity in judging whether an edit is within or outside the scope of the defect, rather than systematic rater disagreement.






Manual labeling surfaces two classes of differences the oracles in RQ1 could not detect: an aggregate gap on root-cause-fix and unrelated-edit rates, and categorical asymmetries in the kinds of repairs each tool produces.
We characterize each below.
RQ3 addresses whether the aggregate gap is statistically significant, using a paired within-bug comparison that has more statistical power than comparing the total counts at this sample size.
RQ2 surfaces what differs between the two tools.
Whether those differences translate into a per-bug preference, and which tool that preference favors, is the question RQ3 answers.

\begin{table}[H]
  \centering
  \small
  \caption{Inter-rater agreement on RQ2 per-patch labels and RQ3 pairwise winner label. \textit{Prev.\ (+)}: positive-class prevalence among non-missing ratings. \textit{Agree}: mean pairwise observed agreement. $\kappa$: Fleiss's chance-corrected agreement.}
  \label{tab:fleiss-kappa}
  \begin{tabular}{@{}llccc@{}}
  \toprule
  \textbf{Label (positive class)} & \textbf{Tool} & \textbf{Prev.\ (+)} & \textbf{Agree} & \textbf{$\kappa$} \\
  \midrule
  \multirow{2}{*}{Patch quality (root cause)}
    & \codex & 0.75 & 0.81 & 0.57 \\
    & \sys   & 0.83 & 0.86 & 0.62 \\
  \midrule
  \multirow{2}{*}{Unrelated change (\texttt{True})}
    & \codex & 0.25 & 0.84 & 0.64 \\
    & \sys   & 0.32 & 0.74 & 0.47 \\
  \midrule
  Winner (pairwise) & --- & --- & 0.68 & 0.54 \\
  \bottomrule
  \end{tabular}
  \end{table}

\subsection{RQ3: Pairwise Patch Quality Analysis}

To answer RQ3, three raters apply the pairwise comparison rubric (Section~\ref{rubric}) to the 144-bug intersection where both \sys and \codex produced a plausible patch, ranking correctness first and comprehensiveness second. 
This paired design removes bug-difficulty noise and gives the test enough statistical power to confirm the quality gap that RQ2 could only suggest directionally.

\begin{figure}[t]
    \centering
    \includegraphics[width=0.7\columnwidth]{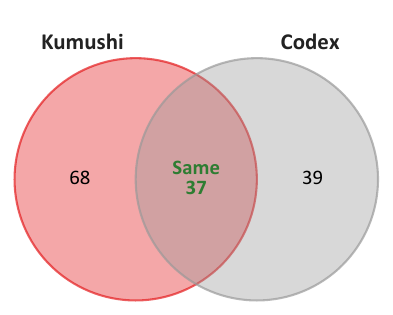}
    \caption{Human preference between \sys and \codex plausible patches. Raters prefer \sys for 68 patches and \codex for 39, with 37 rated equivalent.}
    \label{fig:venn_manual_validation}
\end{figure}

\autoref{fig:venn_manual_validation} summarizes the outcome: \sys is preferred on 68 (47\%), \codex on 39 (27\%), and the two are tied on 37 (26\%). 
Among the 107 decisive bugs where raters reached a clear preference, \sys wins 63.6\% of comparisons; a sign test gives $p = 0.0065$ with a 95\% confidence interval of $(54.4\%, 72.7\%)$ for \sys's win-rate, confirming that the directional gap observed in RQ2 is statistically significant after controlling for bug difficulty.
Inter-rater agreement on the winner label is moderate ($\kappa = 0.54$, Table~\ref{tab:fleiss-kappa}), so the pairwise judgment is reproducible across raters and not driven by a single rater's call.


In the majority of bugs where \sys is preferred, its diversified fault localization surfaces the producer-side defect alongside the ASan-named crash site, allowing \sys to repair the invariant rather than guard the symptom; Cases~A and~B illustrate this pattern.
Case~C illustrates a different dynamic: a \codex-wins case in which the decisive evidence emerged only after synthesizing auxiliary analysis code at repair time, a capability outside \sys's design.


\lstset{
  basicstyle=\ttfamily\small,
  columns=fullflexible,
  keepspaces=true,
  showstringspaces=false,
  captionpos=b,
  aboveskip=4pt,
  belowskip=4pt
}

\subsubsection{Case A: libxml2 Cursor Underflow Across Halt Boundary}
We use Case A as an example to show that \codex misses the root cause location because of missing dynamic information.


The bad state originates in \texttt{xml\allowbreak Halt\allowbreak Parser} at \texttt{parserInternals\allowbreak .c:289}; on a failed buffer grow, it assigns the parser cursor to a one-byte global, \texttt{ctxt->\allowbreak input->\allowbreak cur = BAD\_CAST""}, leaving \texttt{base == cur == end} on a one-byte string.
The mutation that violates the post-halt invariant occurs in \texttt{xmlParse\allowbreak NCName\allowbreak Complex} (\texttt{parser.c:3434}); when its lookahead returns \texttt{c == 0} after a halt, it computes \texttt{ctxt->\allowbreak input->\allowbreak cur -= l} with \texttt{l > cur - base}, underflowing below the global.
ASan reports the subsequent read at \texttt{xmlParseQName} (\texttt{parser.c:8867}), where the caller dereferences the underflowed cursor while testing for \texttt{`:'}.
The producer site, the underflow site, and the crash site each sit in a distinct function.

\begin{figure}[h]
\begin{lstlisting}[style=cstyle,
    caption={Both tools patch \texttt{parser.c:3434}; the asymmetry is in mechanism. \codex{} guards the EOF-state transitions around the underflow site (SF-4 \texttt{corrupt\_state\_tolerance}); \sys{} clamps the cursor arithmetic so the subtraction cannot underflow (RCF-1 \texttt{bound\_check}).},
    label={lst:case-b-fixes}]
// Producer (parserInternals.c:289):
//   xmlHaltParser leaves cur on a 1-byte
//   global -- base == cur == end
  ctxt->input->cur  = BAD_CAST"";
  ctxt->input->base = ctxt->input->cur;
  ctxt->input->end  = ctxt->input->cur;

// Underflow site (xmlParseNCNameComplex):
  if (c == 0) {

// Codex (SF-4, corrupt_state_tolerance):
+   if (ctxt->instate == XML_PARSER_EOF)
+     return(NULL);
    ctxt->input->cur -= l;
    GROW;
+   if (ctxt->instate == XML_PARSER_EOF) {
+     ctxt->input->cur = ctxt->input->base;
+     return(NULL);
+   }

// Kumushi (RCF-1, bound_check):
+   ptrdiff_t back = l;
+   if (ctxt->input->cur
+       - ctxt->input->base < back)
+     back = ctxt->input->cur
+          - ctxt->input->base;
+   ctxt->input->cur -= back;
    GROW;
    if (ctxt->instate == XML_PARSER_EOF)
      return(NULL);
    ctxt->input->cur += back;
    c = CUR_CHAR(l);
  }
\end{lstlisting}
\end{figure}

\sys's Stage-2 ranking surfaced \texttt{xml\allowbreak Halt\allowbreak Parser} as a top FoI on the object-origin tag, with the line \texttt{ctxt->input->cur = BAD\_CAST""} pre-loaded into first-turn context. 
With the post-halt cursor state visible, the agent recognized the invariant \texttt{cur >= base} and patched the underflow site by clamping the subtraction at the cursor extent with \texttt{back = min(l, cur - base)} (RCF-1 \texttt{bound\_check}, Listing~\ref{lst:case-b-fixes}).
The clamp enforces \texttt{cur >= base} at the subtraction site, so the underflow cannot occur whether or not the parser's \texttt{instate} has transitioned to EOF.

\codex's exploration anchored at the ASan-named consumer and walked the call chain through \texttt{xmlParseNCName} into \texttt{xmlParse\allowbreak NCName\allowbreak Complex}. 
\codex{} grep'd the codebase for \texttt{input->\allowbreak cur -=}, found one occurrence at \texttt{parser.c:3434}, and read that as proof the bug lived only there; it never opened \texttt{xml\allowbreak Halt\allowbreak Parser}, even though the ASan report named the producer file as hosting the underflowed buffer.
Without the halt-state fact, \codex framed the bug as an EOF-transition flaw and patched the same line with \texttt{instate} guards around \texttt{GROW}, including a post-\texttt{GROW} reset of \texttt{cur} to \texttt{base} that re-aligns the cursor after the corrupt state has already been created rather than preventing its creation (SF-4 \texttt{corrupt\_\allowbreak state\_\allowbreak tolerance}, Listing~\ref{lst:case-b-fixes}).
The guards fire only when \texttt{instate} has reached EOF, so any halt path that reaches the subtraction before that transition does so unguarded.

Case A shows that two agents can patch the same line under different invariants, with Stage-2 evidence deciding which invariant gets encoded.
This sharpens the localization bottleneck named in observation~(1) of section \ref{sec:preliminary}.
Without dynamic information about the upstream producer, the \codex reaches the patch site but cannot recover the invariant that the producer violates.

\subsubsection{Case B: libsndfile IMA-ADPCM Block Decode}
We use Case B as an example to show that \codex{} reaches the upstream invariant in its reasoning, then drops it after a context-expansion step pulls in irrelevant code.

The defect is a parser-boundary trust violation in the WAV/W64 IMA-ADPCM reader. 
The nibble-unpack loop in \texttt{wavlike\_\allowbreak ima\_\allowbreak decode\_\allowbreak block} assumes its payload is an exact multiple of $4 \times \texttt{channels}$ bytes (eight nibbles per channel per iteration). 
A malformed \texttt{fmt} chunk advertising an inconsistent block-alignment makes the cached block geometry, sample count, and decode loop disagree, so the loop walks past the sample buffer.
The producer is \texttt{ima\_reader\_init}, whose WAV/W64 switch arm already validates \texttt{samplesperblock} against a derived count but does not check that \texttt{blocksize - 4 * channels} is a positive multiple of \texttt{4 * channels}.

\begin{figure}[h]
\begin{lstlisting}[style=cstyle,
    caption={\codex{} states the upstream invariant but does not apply it, guarding writes at the crash site instead (SF-1 \texttt{crash\_site\_guard}). \sys{} encodes the same invariant at the producer (RCF-6 \texttt{invariant\_validation}).},
    label={lst:case-c-fixes}]
// Nibble-unpack: layout invariant
while (blockindx < blocksize) {
  for (chan ...) {
    for (k = 0; k < 4; k++)
      samples[indx] = ...;  // OOB
} }

// Codex (SF-1, crash_site_guard):
+ if (indxstart >= total) break;
  for (chan ...) {
    for (k = 0; k < 4; k++)
+     if (indx < total)
        samples[indx] = ...;
  }

// Kumushi (RCF-6, invariant_validation):
+ if (blockalign < 4 * channels ||
+     (blockalign - 4 * channels)
+      % (4 * channels) != 0)
+   return SFE_WAV_BAD_BLOCKALIGN;
\end{lstlisting}
\end{figure}

\sys{}'s Stage-2 ranking placed \texttt{ima\_reader\_init} at the top of the FoI list, above the ASan-named \texttt{wavlike\_\allowbreak ima\_\allowbreak decode\_\allowbreak block}, contributed by CodeQL's intra-procedural dataflow tag and the dynamic call trace; the surrounding switch arm and the existing \texttt{samplesperblock} guard were pre-loaded into first-turn context. 
The agent extended the same arm with the missing divisibility precondition, returning \texttt{SFE\_\allowbreak WAV\_\allowbreak BAD\_\allowbreak BLOCKALIGN} when the input violates the nibble-layout invariant (RCF-6 \texttt{invariant\_validation}, Listing~\ref{lst:case-c-fixes}).

\codex{}'s trace identifies the same invariant and then drops it.
After reading the decode loop and the init function, \codex{} stated the upstream check explicitly, that \texttt{(blockalign - 4*channels)} should be divisible by \texttt{(4*channels)}.
After a call-graph sweep on \texttt{decode\_block} broadened context with the encoder, the read-block helper, and the seek paths, \codex{} dropped the divisibility precondition in favor of in-loop bounds clamping.
Only the floor check \texttt{blocksize < 4 * channels} survived into the emitted patch, which guards each indexed write at the consumer site rather than rejecting the malformed geometry upstream (SF-1 \texttt{crash\_site\_guard}, Listing~\ref{lst:case-c-fixes}).

Case B instantiates observation~(2) of section \ref{sec:preliminary}. 
When unfiltered context expands during exploration, even a correctly identified upstream check can be displaced by off-path code patterns.
\sys{}'s Stage-2 ranking places the upstream initializer at the agent's first read, so the call-graph drift that displaces the right invariant in \codex{}'s exploration never has the chance to occur.

\subsubsection{Case C: mruby Heap Buffer Overflow}
We use Case C as an example to show that a defect that lies outside \sys{}'s reach.
The root cause resides in the compiler's bytecode generator, separated from the crash by thousands of frames in the call graph.
This places it outside the scope of \sys{}'s static analysis and renders it indistinguishable from normal behavior under dynamic analysis.

The proof of concept \verb|~(:regex or 1*1)| triggers a heap-buffer-overflow at \texttt{vm.c\allowbreak :1120}, where the VM indexes a one-entry literal pool with \texttt{b = 1}.
The crash site, however, is not the defect site.
The underlying bug is in \texttt{gen\_\allowbreak uniop} (\texttt{codegen.c:1137}), the only unary peephole rewrite path that lacks a \texttt{no\_peephole()} guard.
As Listing~\ref{lst:bytecode-drift} shows, the failing case rewrites \texttt{OP\_LOADI\_1} (2 bytes) into \texttt{OP\_LOADINEG} (3 bytes) after an \texttt{OP\_JMPIF} target has already been patched.
The jump target therefore remains at the old byte offset and lands in the middle of the rewritten instruction, so the VM decodes operand byte \texttt{0x02} as a fresh opcode and eventually reads past the literal pool.

\begin{figure}[h]
\begin{lstlisting}[style=cstyle,
    caption={Bytecode drift after peephole rewrite.},
    label={lst:bytecode-drift}]
before rewrite:
  2: OP_JMPIF    r3, +8    // target = 10
  8: OP_LOADI_1  r3        // 2 bytes
 10: <next instruction>
after gen_uniop folds ~1 to -2:
  2: OP_JMPIF    r3, +8    // still targets 10
  8: OP_LOADINEG r3, 2     // 3 bytes
 10: 0x02                  // jump lands in operand byte
\end{lstlisting}
\end{figure}

\begin{figure}[h]
\begin{lstlisting}[style=cstyle,
    caption={Minimal guard for unary peephole folding.},
    label={lst:uniop-fix}]
static mrb_bool
gen_uniop(codegen_scope *s, mrb_sym sym, uint16_t dst)
{
  if (no_peephole(s)) return FALSE;
  struct mrb_insn_data data = mrb_last_insn(s);
  ...
}
\end{lstlisting}
\end{figure}

\texttt{gen\_\allowbreak uniop} runs during compilation, hundreds of frames before \texttt{vm.c\allowbreak :1120}, and the bytecode buffer that links compiler to VM is opaque to static dataflow.
\sys{}'s evidence sources score candidates by uniqueness in the trace and connectivity through dataflow tags, but no dataflow edge leads back to the rewriter and bytecode-generation routines appear throughout the trace, so neither signal surfaces it.
\sys{} therefore cannot localize the root cause and emits a patch that validates the decoded literal-pool index \texttt{b} before the VM reads \texttt{pool[b]} (SF-1 \texttt{crash\_site\_guard}).
The patch suppresses the observed crash but does not repair the violated compiler invariant.

The \codex{} run, by contrast, generated a bytecode disassembler during repair, exposed the one-byte drift in Listing~\ref{lst:bytecode-drift}, and traced the problem back to a missing \texttt{no\_peephole()} condition on the unary folding path. 
Its patch inserts the guard at the caller-side fold site in \texttt{codegen.c}, blocking the same unguarded fold path as the ground-truth fix, which places the check inside \texttt{gen\_\allowbreak uniop} (RCF-4 \texttt{control\_flow\_fix}, Listing~\ref{lst:uniop-fix}).

Case C demonstrates that crash-anchored ranking misses defects whose decisive evidence emerges only at repair time.
\codex{} reached the rewriter because a general-purpose coding agent can construct a custom program-analysis tool when its reasoning requires one.
\sys{}'s current Stage-2 evidence sources have no edge to follow back across the encoded-bytecode boundary, so this class of bugs is unreachable by the present pipeline.

\smallskip
\noindent\textit{\textbf{Takeaway.}}
In summary, the three cases identify the factors that drive the quality gap.
\sys{}'s evidence-weighted ranking surfaces upstream producers that the ASan stack does not name (Case~A) and places them at the agent's first read, preempting the call-graph drift that displaces the right invariant during \codex{}'s exploration (Case~B).
Both moves push patches toward root-cause categories (RCF-1 \texttt{bound\_check}, RCF-6 \texttt{invariant\_validation}) rather than crash-site or downstream guards.
Root-cause-anchored patching's strength comes from grounding the agent in evidence the LLM cannot recover by reading code alone.
Its limitation appears when the decisive evidence sits behind a boundary that no static or dynamic source can cross (Case~C), in which case a general-purpose agent that synthesizes its own analysis tool can outperform it.

\ignore{\begin{table}[H]
  \centering
  \small
  \caption{Design and test by RQ.}
  \label{tab:rq-arc}
  \begin{tabular}{@{}llll@{}}
  \toprule
  \textbf{Section} & \textbf{Design} & \textbf{Test} & \textbf{$p$-value} \\
  \midrule
  RQ1 & Unpaired marginal  & Fisher's exact & $p = 1.0$ \\
  RQ2 & Unpaired aggregate & Fisher's exact & $p = 0.14$, $0.19$ \\
  RQ3 & Paired within-bug  & Sign test      & $p = 0.0065$ \\
  \bottomrule
  \end{tabular}
  \end{table}
The same gap appears at different resolutions across the three RQs (\autoref{tab:rq-arc}).
Oracle-based marginal counts cannot separate frontier tools (RQ1); unpaired aggregate labels are directionally consistent but underpowered at this sample size (RQ2); a paired within-bug comparison removes bug-difficulty noise and detects the difference at $p = 0.0065$ (RQ3).
}

\section{Related Work}
\smallskip
\noindent\textit{\textbf{Automated Program Repair.}}
Automated program repair (APR) has been extensively studied for generating patches that transform defective or vulnerable programs into correct or secure ones. Existing APR techniques can be broadly categorized into search-based, template-based, semantic-based, and learning-based approaches.

\textit{Search-based repair.}
Search-based approaches formulate patch generation as a search problem over a space of program variants~\cite{Goues2012GenProgAG,Gao2019CrashavoidingPR}. GenProg~\cite{Goues2012GenProgAG}, for example, generates candidate variants through mutation and crossover and evaluates them against test suites. Fix2Fit~\cite{Gao2019CrashavoidingPR} further incorporates fuzzing to refine the search space and guide the generation of crash-avoiding patches.

\textit{Template-based repair.}
Template-based approaches rely on predefined patch patterns or repair templates to produce candidate fixes~\cite{Gao2016BovInspectorAI,Li2022RegexScalpelRE,Son2013FixMU,Xu2019VFixVP,Kim2013AutomaticPG}. BovInspector~\cite{Gao2016BovInspectorAI} uses safe API templates to repair buffer-overflow vulnerabilities. RegexScalpel~\cite{Li2022RegexScalpelRE} employs regular-expression templates to mitigate ReDoS vulnerabilities. VFix~\cite{Xu2019VFixVP} uses two sets of repair templates to address null-pointer-dereference vulnerabilities.

\textit{Semantic-based repair.}
Semantic-based approaches exploit program semantics, vulnerability properties, or inferred constraints to synthesize patches~\cite{Hong2020SAVERSP,Tonder2018StaticAP,gao2021beyond,Huang2019UsingSP,Zhang2022ProgramVR,Gao2015SafeMF,Shariffdeen2021ConcolicPR}. SAVER~\cite{Hong2020SAVERSP} constructs constraints from variable states and control-/data-flow graphs to generate repairs. ExtractFix~\cite{gao2021beyond} derives crash-free constraints and uses them to synthesize candidate patches. VulnFix~\cite{Zhang2022ProgramVR} infers program invariants and leverages them to guide vulnerability repair.

\textit{Learning-based repair.}
More recently, learning-based APR techniques have used neural models and large language models to translate vulnerable code into repaired code~\cite{Fu2023VisionTI,Harer2018LearningTR,Chen2021NeuralTL,Fu2022VulRepairAT,Zhou2024OutOS}. VulRepair~\cite{Fu2022VulRepairAT}, for instance, fine-tunes CodeT5~\cite{Wang2021CodeT5IU} for vulnerability repair. Other work has explored prompting strategies that improve the ability of LLMs to generate security patches~\cite{Pearce2021ExaminingZV,Nong2025APPATCHAA,Kim2025LogsIP}.

\smallskip
\noindent\textit{\textbf{Root Cause Analysis.}}
Root cause analysis aims to identify the program locations and execution conditions responsible for a failure or vulnerability. Prior work has shown that combining static and dynamic analysis can improve the localization of root causes and provide useful guidance for patch generation~\cite{Xu2019VFixVP}. Existing techniques diagnose bugs and vulnerabilities using statistical, trace-based, and semantic information.

\textit{Statistical root cause analysis.}
Statistical techniques rank candidate program locations according to their correlation with failing executions~\cite{Blazytko2020AURORASC,Park2024BenzeneAP,Ochiai,Tarantula}. AURORA~\cite{Blazytko2020AURORASC} compares traces from crashing and non-crashing inputs and uses statistical analysis to identify likely root-cause locations. Such techniques are effective when sufficient execution evidence is available, but their accuracy depends on the quality and diversity of the collected traces and may degrade when failures are triggered by complex semantic conditions.

\textit{Semantic root cause analysis.}
Semantic techniques use program reasoning to explain how a proof-of-concept input reaches a vulnerable state~\cite{Yagemann2021ARCUSSR,Yagemann2021AutomatedBH}. ARCUS~\cite{Yagemann2021ARCUSSR}, for example, builds on symbolic execution with angr~\cite{shoshitaishvili2016state} to trace a vulnerability and identify its root cause. Compared with statistical techniques, semantic approaches can provide more precise explanations of vulnerability-triggering conditions, but they often depend on heavyweight program analyses and may face scalability challenges on complex programs.





\section{Discussion}


\smallskip
\noindent\textit{\textbf{Oracle-Based Evaluation Is No Longer Sufficient.}}
The automated vulnerability repair community has long acknowledged that replay-based oracles are an imperfect proxy for patch quality\cite{Hu2025SoKAV}. A patch may suppress the observed failure while leaving the underlying defect intact. The community recognized this limitation but tolerated it. When the field's best tools produced meaningfully different oracle-passing rates, oracle-based evaluation could still discriminate between them. \patchagent, \looprepair, and earlier prompt-based approaches produce meaningfully different plausible patch rates (71.9\%, 2.2\%, respectively), and oracle metrics alone are sufficient to rank them.

We find that oracle-based evaluation no longer separates the latest LLM-based agents. Our evaluation shows that \sys, a specialized root-cause-driven agent, and \codex, a general-purpose coding agent, are statistically indistinguishable under all three oracles, yet expert assessment reveals a significant quality gap in favor of \sys ($p = 0.0065$). This is not an artifact of our benchmark or our tools. It reflects a structural shift in agent capability. General-purpose coding agents now reliably produce oracle-passing patches across a wide range of vulnerability types, saturating the metric the community has long relied on.

The implication for the APR community is direct. As coding agents continue to improve, oracle-passing rate will fail to separate tools that produce root-cause fixes from those that produce symptom patches. 
Therefore, they require evaluation methods capable of assessing \emph{what} a patch fixes, not \emph{whether} the observed crash is suppressed. 
The two-tier metric introduced in this work is one step in that direction, pairing automated oracles with structured expert assessment of patch comprehensiveness. 
We hope future work can build automated proxies for expert judgment that can flag symptom patches before they are counted as successes.


\smallskip
\noindent\textit{\textbf{Limitations of Crash-Anchored Localization.}}
\sys's root-cause-driven design assumes the defect site is reachable from the crash site through static or dynamic evidence. This assumption holds for the majority of bugs in our benchmark, where the producer, the underflow site, and the crash site are connected by dataflow edges or execution traces that \hyperref[sec:design:fl]{Stage~1} can collect and \hyperref[sec:design:rank]{Stage~2} can rank. Case~C shows where it breaks down. When the defect site is separated from the crash by an opaque boundary, such as an encoded bytecode buffer or an inter-process channel, no static dataflow edge and no dynamic trace leads back to it. Evidence-weighted ranking cannot surface what no evidence source can reach.

This is a structural limit of crash-anchored localization, not an implementation gap specific to \sys. 
Any approach that grounds localization in crash-site evidence will face the same issue.
Case~C also illustrates what can fill the gap. 
A general-purpose agent that synthesizes its own analysis tools at repair time can cross boundaries that pre-computed evidence sources cannot. 
The tradeoff is that such agents operate without the focused context that evidence-weighted ranking provides, making them more susceptible to the drift pattern that motivates \sys's design in the first place.

Two additional scope constraints bound the current prototype. 
First, the quality of \hyperref[sec:design:fl]{Stage~1}'s candidate pool depends on how many distinct crash paths the fuzzer finds within the time budget. Bugs with narrow trigger conditions may not diversify enough to expose root-cause-correlated functions beyond the crash stack. 
Second, the pipeline currently assumes sanitizer instrumentation and fuzzing infrastructure, which limits direct applicability to languages and ecosystems where such tooling is unavailable. 
We leave extending \sys to them to future work.


\smallskip
\noindent\textit{\textbf{Toward Automated Patch Quality Validation.}}
Manual expert assessment, as applied in our tier-2 evaluation, does not scale to the patch volumes that LLM-based repair pipelines now produce. 
Our evaluation required approximately 16 minutes per patch~\cite{Tao2014AutomaticallyGP}, and even with eight raters working in parallel, coverage was bounded by the 144-bug intersection where both tools produced a plausible patch. As repair systems improve and benchmarks grow, this bottleneck will become more acute.

Our taxonomy of symptom-fix strategies in Table~\ref{tab:symptom-taxonomy} suggests a path forward. 
The nine failure modes we identify are not arbitrary. Crash-site guards, downstream guards, and trigger blocks share a structural signature in which the patch modifies code at or near the ASan-reported crash site without touching any function that the evidence-weighted ranking would have flagged as a root-cause candidate. Value masking and corrupt-state tolerance leave the producer intact while suppressing its observable effect. 
These structural fingerprints are in principle detectable without human review. An automated checker can compare the patch's edit locations against the FoI ranking, determine whether the modified code touches a producer or only a consumer, and flag patches whose diffs are confined to the crash stack. We hope future work uses this taxonomy as a foundation for automated validators that can triage symptom patches before they reach a human reviewer, closing the scalability gap that tier-2 assessment currently cannot bridge.

\section{Conclusion}
This work introduces \sys, a vulnerability patching system that addresses two limitations of existing LLM-based tools: (1) static-only fault localization that fails to pinpoint root causes, and (2) unfiltered context that drives agents toward crash-site suppression instead of root-cause repair.
In particular, \sys uses two innovative techniques: diversified fault localization and evidence-weighted context ranking.
Based on our evaluation of patching 178 C/C++ vulnerabilities, \sys generates plausible patches for over 90\% of vulnerabilities under the replay-based oracle and produces more high-quality patches than commercial general-purpose coding agents under blind human expert assessment, which outperforms the state-of-the-art automatic root cause repair techniques in both academia and industry.


\bibliographystyle{ACM-Reference-Format}
\bibliography{biblio}

\appendix
\section{Repair Strategies Definition}
\label{app:repair_strategies}
The definition of root cause fix strategies is in \autoref{tab:fix-mechanisms}.
The definition of symptom fix strategies is in \autoref{tab:failure-modes}.

\begin{table*}[htbp]
\begin{minipage}[t]{0.49\textwidth}
\centering
\small
\begin{tabular}{@{}lp{0.65\linewidth}@{}}
\toprule
\textbf{Mechanism} & \textbf{Definition} \\
\midrule
\texttt{bound\_check} &
  Enforces the correct index, cursor, remaining-length, or object
  extent before access (sizes, lengths, indices, offsets vs.\
  capacities). \\
\texttt{api\_contract} &
  Enforces the intended boundary contract for nullable returns,
  EOF/error propagation, output parameters, string termination, or
  memory API semantics. \\
\texttt{size\_arithmetic} &
  Repairs overflow, underflow, divide-by-zero, capacity accounting,
  or exact bytes-needed calculations before the result is trusted. \\
\texttt{control\_flow\_fix} &
  Repairs the predicate, iteration, recursion, parser transition, or
  error-path branch that produces the bad state. \\
\texttt{ownership\_repair} &
  Repairs the ownership or lifetime rule that creates stale,
  double-freed, or dangling state (UAF, double-free, freed-pointer
  reuse). \\
\texttt{invariant\_validation} &
  Rejects or normalises invalid input at the producer/parse boundary
  before it becomes trusted internal state (non-zero divisor,
  plausible count, valid enum). \\
\texttt{state\_sync} &
  Keeps derived state, cached metadata, counters, lengths, or
  helper-visible buffer extents synchronised after mutation. \\
\texttt{type\_representation} &
  Repairs signedness, integer width, truncation, or producer/consumer
  type-domain agreement. \\
\bottomrule
\end{tabular}
\captionof{table}{Root-cause fix Strategies}
\label{tab:fix-mechanisms}
\end{minipage}%
\hfill
\begin{minipage}[t]{0.49\textwidth}
\centering
\small
\begin{tabular}{@{}lp{0.6\linewidth}@{}}
\toprule
\textbf{Failure mode} & \textbf{Definition} \\
\midrule
\texttt{crash\_site\_guard} &
  Immediate guard around the observed dereference, division, shift,
  copy, or read while the upstream invalid state remains intact. \\
\texttt{downstream\_guard} &
  Patches a caller or consumer while leaving the producer or shared
  helper defect intact, so equivalent callers still trigger the same
  bug. \\
\texttt{value\_masking} &
  Hides the bad value by clamping, replacing, padding, inserting a
  dummy object, or forcing a sentinel/default. \\
\texttt{corrupt\_state\_tolerance} &
  Detects or tolerates corrupted state after it has already been
  created instead of preventing creation. \\
\texttt{trigger\_block} &
  Blocks one observed input shape (the proof-of-concept) without
  enforcing the general invariant. \\
\texttt{incomplete\_coverage} &
  Fixes one crash location while leaving another path for the same
  root cause unpatched. \\
\texttt{lifetime\_masking} &
  Pins, retains, or avoids freeing an object to suppress a lifetime
  symptom while leaving the ownership transition wrong. \\
\texttt{wrong\_location} &
  Modifies code that is not on the causal path, or addresses a
  different bug than the one reported. \\
\texttt{ineffective\_guard} &
  Adds a check that cannot fire, checks the wrong condition, or
  enforces a nearby but non-causal invariant. \\
\bottomrule
\end{tabular}
\captionof{table}{Symptom-fix Strategies}
\label{tab:failure-modes}
\end{minipage}
\end{table*}

\section{Benchmark}
\label{app:bench}
\autoref{tab:bugs} shows all the bugs in our dataset.

\begin{table*}[t]
\centering
\caption{Benchmark}
\label{tab:bugs}
\small 
\renewcommand{\arraystretch}{1.05}
\resizebox{\textwidth}{!}{
\begin{tabular}{@{}lll|lll|lll@{}}
\toprule
\textbf{Project} & \textbf{Commit} & \textbf{Bug Type} &
\textbf{Project} & \textbf{Commit} & \textbf{Bug Type} &
\textbf{Project} & \textbf{Commit} & \textbf{Bug Type}\\
\midrule
assimp        & \texttt{0422dff} & heap\_buffer\_overflow     & gpac      & \texttt{a6b6408} & heap\_out\_of\_bound       & md4c          & \texttt{3478ec6} & heap\_buffer\_overflow\_a\\
assimp        & \texttt{2d44861} & unknown\_write             & gpac      & \texttt{a8bc2c8} & heap\_out\_of\_bound       & md4c          & \texttt{3478ec6} & heap\_buffer\_overflow\_b\\
assimp        & \texttt{565539b} & heap\_buffer\_overflow     & gpac      & \texttt{b1042c3} & null\_pointer\_deref       & md4c          & \texttt{db9ab41} & heap\_buffer\_overflow\\
c-blosc       & \texttt{01df770} & heap\_buffer\_overflow     & gpac      & \texttt{b6b6360} & null\_pointer\_deref       & md4c          & \texttt{db9ab41} & unknown\_read\\
c-blosc       & \texttt{41f3a2e} & heap\_buffer\_overflow     & gpac      & \texttt{bb9ee4c} & heap\_out\_of\_bound       & mruby         & \texttt{0ed3fcf} & heap\_out\_of\_bound\\
c-blosc2      & \texttt{38b23d5} & negative\_size\_param      & gpac      & \texttt{be23476} & heap\_out\_of\_bound       & mruby         & \texttt{4c196db} & heap\_out\_of\_bound\\
c-blosc2      & \texttt{4f6d42a} & invalid\_free              & gpac      & \texttt{ca1b48f} & heap\_out\_of\_bound       & mruby         & \texttt{55b5261} & null\_pointer\_deref\\
c-blosc2      & \texttt{6fc4790} & heap\_buffer\_overflow\_b  & gpac      & \texttt{cc95b16} & heap\_out\_of\_bound       & mruby         & \texttt{8aec568} & use\_after\_free\\
c-blosc2      & \texttt{81c2fcd} & heap\_buffer\_overflow\_b  & gpac      & \texttt{d2de8b5} & stack\_out\_of\_bound      & mruby         & \texttt{af5acf3} & use\_after\_free\\
c-blosc2      & \texttt{81c2fcd} & heap\_buffer\_overflow\_c  & gpac      & \texttt{de7f3a8} & heap\_out\_of\_bound       & mruby         & \texttt{b4168c9} & use\_after\_free\\
c-blosc2      & \texttt{aebf2b9} & heap\_buffer\_overflow     & gpac      & \texttt{ebedc7a} & use\_after\_free           & mruby         & \texttt{bdc244e} & null\_pointer\_deref\\
c-blosc2      & \texttt{cb15f1b} & unknown\_read              & gpac      & \texttt{fc9e290} & heap\_out\_of\_bound       & mruby         & \texttt{bf5bbf0} & use\_after\_free\\
binutils      & \texttt{515f23e} & divide\_by\_zero           & h3        & \texttt{f581626} & global\_buffer\_overflow   & mruby         & \texttt{c30e6eb} & heap\_out\_of\_bound\\
binutils      & \texttt{c48935d} & heap\_buffer\_overflow     & hoextdown & \texttt{933f9da} & heap\_buffer\_overflow     & mruby         & \texttt{d1f1b4e} & null\_pointe\_deref\\
coreutils     & \texttt{658529a} & heap\_buffer\_overflow     & hostap    & \texttt{703c2b6} & heap\_buffer\_overflow     & openexr       & \texttt{115e42e} & heap\_buffer\_overflow\_a\\
coreutils     & \texttt{68c5eec} & negative\_size\_param      & hostap    & \texttt{8112131} & unknown\_read              & openexr       & \texttt{115e42e} & heap\_buffer\_overflow\_b\\
coreutils     & \texttt{8d34b45} & memcpy\_param-overlap      & hostap    & \texttt{a6ed414} & heap\_buffer\_overflow\_a  & openexr       & \texttt{7c40603} & stack\_buffer\_overflow\\
coreutils     & \texttt{ca99c52} & global\_buffer\_overflow   & hostap    & \texttt{a6ed414} & heap\_buffer\_overflow\_b  & radare2       & \texttt{0927ed3} & heap\_out\_of\_bound\\
jasper        & \texttt{3c55b39} & heap\_buffer\_overflow     & htslib    & \texttt{dd6f0b7} & unknown\_read              & radare2       & \texttt{0be8f25} & heap\_out\_of\_bound\\
jasper        & \texttt{b9be3d9} & integer\_overflow          & hunspell  & \texttt{1c1f34f} & heap\_buffer\_overflow     & radare2       & \texttt{108dc76} & heap\_out\_of\_bound\\
libjpeg-turbo & \texttt{208d927} & heap\_buffer\_overflow     & hunspell  & \texttt{473241e} & heap\_buffer\_overflow     & radare2       & \texttt{27fe803} & null\_pointer\_deref\\
libjpeg-turbo & \texttt{3212005} & null\_ptr\_dereference     & hunspell  & \texttt{6291cac} & heap\_buffer\_overflow\_a  & radare2       & \texttt{4823451} & heap\_out\_of\_bound\\
libjpeg-turbo & \texttt{4f24016} & stack\_buffer\_overflow    & hunspell  & \texttt{6291cac} & heap\_buffer\_overflow\_b  & radare2       & \texttt{4b22fc5} & null\_pointer\_deref\\
libjpeg-turbo & \texttt{ae8cdf5} & heap\_buffer\_overflow     & hunspell  & \texttt{6291cac} & stack\_buffer\_overflow\_a & radare2       & \texttt{515e592} & heap\_out\_of\_bound\\
libtiff       & \texttt{0ba5d88} & heap\_buffer\_overflow     & hunspell  & \texttt{6291cac} & stack\_buffer\_overflow\_b & radare2       & \texttt{5c0bde8} & null\_pointer\_deref\\
libtiff       & \texttt{2c00d31} & divide\_by\_zero           & hunspell  & \texttt{74b08bf} & heap\_buffer\_overflow\_a  & radare2       & \texttt{72ffc02} & null\_pointer\_deref\\
libtiff       & \texttt{3144e57} & invalid\_shift             & hunspell  & \texttt{74b08bf} & heap\_buffer\_overflow\_b  & radare2       & \texttt{7cfd367} & use\_after\_free\\
libtiff       & \texttt{6a984bf} & heap\_buffer\_overflow     & hunspell  & \texttt{74b08bf} & heap\_double\_free         & radare2       & \texttt{95b648f} & heap\_out\_of\_bound\\
libtiff       & \texttt{891b1b9} & heap\_buffer\_overflow     & hunspell  & \texttt{82b9212} & heap\_use\_after\_free     & radare2       & \texttt{9650e3c} & use\_after\_free\\
libtiff       & \texttt{8b6e80f} & heap\_buffer\_overflow     & hunspell  & \texttt{ddec95b} & heap\_buffer\_overflow     & radare2       & \texttt{9bcc98f} & heap\_out\_of\_bound\\
libtiff       & \texttt{9a72a69} & divide\_by\_zero           & irssi     & \texttt{afcb483} & heap\_use\_after\_free     & radare2       & \texttt{a16cb20} & heap\_out\_of\_bound\\
libtiff       & \texttt{acb5bb9} & heap\_buffer\_overflow     & irssi     & \texttt{b472570} & heap\_buffer\_overflow\_a  & radare2       & \texttt{a58b8d4} & heap\_out\_of\_bound\\
libtiff       & \texttt{c421b99} & heap\_buffer\_overflow     & irssi     & \texttt{b472570} & heap\_buffer\_overflow\_b  & radare2       & \texttt{cf780fd} & use\_after\_free\\
libtiff       & \texttt{e596d4e} & divide-by-zero             & krb5      & \texttt{d864d74} & heap\_buffer\_overflow     & radare2       & \texttt{d026503} & heap\_out\_of\_bound\\
libtiff       & \texttt{f3069a5} & heap\_buffer\_overflow     & libmobi   & \texttt{1297ee0} & heap\_out\_of\_bound       & radare2       & \texttt{d17a7bd} & use\_after\_free\\
libxml2       & \texttt{362b322} & null\_ptr\_dereference     & libmobi   & \texttt{4b60805} & heap\_out\_of\_bound       & radare2       & \texttt{d22d160} & use\_after\_free\\
libxml2       & \texttt{4ea74a4} & heap\_buffer\_overflow     & libmobi   & \texttt{afa8ce1} & stack\_out\_of\_bound      & radare2       & \texttt{eca58ce} & heap\_out\_of\_bound\\
libxml2       & \texttt{8f30bdf} & heap\_buffer\_overflow     & libmobi   & \texttt{eafc415} & null\_pointer\_deref       & sleuthkit     & \texttt{34f995d} & heap\_buffer\_overflow\_a\\
libxml2       & \texttt{cbb2716} & heap\_buffer\_overflow     & libmobi   & \texttt{eb4a262} & null\_pointer\_deref       & sleuthkit     & \texttt{34f995d} & heap\_buffer\_overflow\_b\\
libxml2       & \texttt{db07dd6} & heap\_buffer\_overflow     & libplist  & \texttt{491a3ac} & heap\_buffer\_overflow\_a  & sleuthkit     & \texttt{38a13f9} & heap\_buffer\_overflow\\
gpac          & \texttt{0b29a41} & use\_after\_free           & libplist  & \texttt{491a3ac} & heap\_buffer\_overflow\_b  & sleuthkit     & \texttt{82d254b} & heap\_buffer\_overflow\\
gpac          & \texttt{112767e} & heap\_out\_of\_bound       & libplist  & \texttt{491a3ac} & heap\_use\_after\_free     & sleuthkit     & \texttt{d9b19e1} & heap\_buffer\_overflow\\
gpac          & \texttt{1b77837} & heap\_out\_of\_bound       & libsndfile& \texttt{2b4cc4b} & heap\_buffer\_overflow     & wasm3         & \texttt{139076a} & use\_after\_free\\
gpac          & \texttt{3ffe59c} & heap\_out\_of\_bound       & libsndfile& \texttt{4819cad} & heap\_buffer\_overflow     & wasm3-harness & \texttt{0124fd5} & heap\_buffer\_overflow\\
gpac          & \texttt{4607052} & double\_free               & libsndfile& \texttt{932aead} & negative\_size\_param      & wasm3-harness & \texttt{355285d} & global\_buffer\_overflow\\
gpac          & \texttt{4925c40} & use\_after\_free           & libsndfile& \texttt{b706e62} & heap\_buffer\_overflow     & wasm3-harness & \texttt{4f0b769} & heap\_use\_after\_free\\
gpac          & \texttt{49cb88a} & heap\_out\_of\_bound       & libsndfile& \texttt{fe49327} & heap\_buffer\_overflow     & wasm3-harness & \texttt{4f0b769} & unknown\_read\\
gpac          & \texttt{4c77303} & heap\_out\_of\_bound       & libtpms   & \texttt{e563166} & stack\_buffer\_overflow    & wasm3-harness & \texttt{4f0b769} & unknown\_write\\
gpac          & \texttt{50a60b0} & null\_pointer\_deref       & libxml2   & \texttt{20f5c73} & global\_buffer\_overflow   & wasm3-harness & \texttt{bc32ee0} & segv\_on\_unknown\_address\\
gpac          & \texttt{514a3af} & stack\_out\_of\_bound      & libxml2   & \texttt{5f4ec41} & global\_buffer\_overflow   & wasm3-harness & \texttt{bc32ee0} & unknown\_read\\
gpac          & \texttt{6f28c4c} & null\_pointer\_deref       & libxml2   & \texttt{7fbd454} & global\_buffer\_overflow   & yasm          & \texttt{84be2ee} & heap\_out\_of\_bound\\
gpac          & \texttt{78f5269} & heap\_out\_of\_bound       & libxml2   & \texttt{9ef2a9a} & global\_buffer\_overflow\_a& yasm          & \texttt{9defefa} & heap\_out\_of\_bound\\
gpac          & \texttt{7a6f636} & stack\_out\_of\_bound      & libxml2   & \texttt{9ef2a9a} & global\_buffer\_overflow\_b& yasm          & \texttt{ecb47f1} & null\_pointer\_deref\\
gpac          & \texttt{7e2cb01} & use\_after\_free           & libxml2   & \texttt{b167c73} & global\_buffer\_overflow\_a& zstd          & \texttt{0a96d00} & heap\_buffer\_overflow\\
gpac          & \texttt{7e2e92f} & use\_after\_free           & libxml2   & \texttt{b167c73} & global\_buffer\_overflow\_b& zstd          & \texttt{2fabd37} & global\_buffer\_overflow\\
gpac          & \texttt{7edc40f} & null\_pointer\_deref       & libxml2   & \texttt{b167c73} & global\_buffer\_overflow\_c& zstd          & \texttt{3cac061} & heap\_buffer\_overflow\\
gpac          & \texttt{89a80ca} & heap\_out\_of\_bound       & libxml2   & \texttt{b167c73} & global\_buffer\_overflow\_d& zstd          & \texttt{6f40571} & unknown\_read\\
gpac          & \texttt{8db20cb} & heap\_out\_of\_bound       & libxml2   & \texttt{b167c73} & global\_buffer\_overflow\_e& zstd          & \texttt{d68aa19} & heap\_buffer\_overflow\\
gpac          & \texttt{8f3088b} & null\_pointer\_deref       & lz4       & \texttt{7654a5a} & heap\_buffer\_overflow     &               &                  & \\
gpac          & \texttt{94cf5b1} & stack\_out\_of\_bound      & lz4       & \texttt{9d20cd5} & invalid\_free              &               &                  & \\
\bottomrule
\end{tabular}
}
\end{table*}

\lstdefinestyle{prompt}{
  basicstyle=\ttfamily\footnotesize,
  breaklines=true,
  breakatwhitespace=true,
  breakindent=0pt,
  columns=fullflexible,
  keepspaces=true,
  showstringspaces=false,
  upquote=true,
  frame=single,
  framesep=4pt,
  rulecolor=\color{black!40},
  backgroundcolor=\color{black!3},
  linewidth=\textwidth,
  resetmargins=true,
  xleftmargin=0pt,
  xrightmargin=0pt,
  aboveskip=6pt,
  belowskip=6pt,
  literate=%
    {`}{{\textasciigrave}}1
    {~}{{\textasciitilde}}1
    {\^}{{\textasciicircum}}1,
}

\clearpage
\onecolumn
\section{LLM System Prompt}
\label{appendix:llm-prompt}

This appendix reproduces verbatim the system prompt sent to the LLM
patch agent. The two placeholders \texttt{\{crash\_report\}} and
\texttt{\{rca\_summary\}} are substituted at run time with the
sanitizer report and the ranked Root~Cause~Analysis summary,
respectively.

\begin{lstlisting}[style=prompt,caption={System prompt template used by the \sys.},label={lst:system-prompt}]
You are an expert C/C++ security researcher and patch developer. Your job is to analyze a vulnerability in a C/C++ project, understand its root cause, and develop a minimal, correct patch that fixes the vulnerability without breaking existing functionality.

## Crash Report (ASan/Sanitizer Output)

{crash_report}

## Root Cause Analysis Summary

{rca_summary}

## How the RCA Pipeline Works

The Root Cause Analysis (RCA) results you see were produced by a multi-stage pipeline that combines static analysis and dynamic tracing to rank candidate functions by their likelihood of containing the bug:

1. **Stack Trace Analysis** (source: STACK_TRACE) -- Parses the ASAN/sanitizer crash report to extract the exact crash location and call stack. These are the highest-confidence candidates -- they are directly on the crash path.

2. **Dynamic Call Tracing** (source: CALL_TRACE) -- Instruments the binary and replays the crashing input to record every function called at runtime. Functions are ranked by their distance from the crash site (closer = more relevant) and call frequency.

3. **Static Variable Dependencies** (source: VAR_DEP) -- Performs static data-flow analysis to find functions that handle data flowing toward the crash location. These may reveal upstream allocation, sizing, or validation functions that are the true root cause.

4. **Fuzzing Coverage Analysis** (source: AURORA) -- Uses fuzzer corpus and crash coverage to score functions by how strongly their coverage correlates with crashes.

Each FOI (Function of Interest) cluster groups related functions from these analyses. Clusters are ranked by a combination of how many independent analyses flagged them and the priority of those sources (STACK_TRACE is highest priority; AURORA varies by score -- check each cluster's reasoning for its confidence level; then VAR_DEP, then CALL_TRACE). A function flagged by multiple analyses ranks higher than one flagged by a single high-priority source. The top-ranked clusters are the most likely to contain the root cause, but the actual buggy code may be in a caller, a callee, or a related helper -- not necessarily the crash site itself.

## Workflow

### Phase 1: Deep Exploration (do NOT edit code yet)
Before making any edits, you MUST thoroughly understand the vulnerability by examining the relationship between the RCA candidates and the bug:

1. **Study the RCA clusters**: The most relevant clusters are shown above with full source code -- review them first. Use `get_rca_results(index)` to view source for any additional clusters listed compactly. Pay attention to the source of each cluster -- clusters flagged by multiple analyses deserve extra attention.

2. **Understand the crash mechanism**: From the crash report, identify the bug type (buffer overflow, use-after-free, null dereference, etc.). Identify the exact memory operation that fails and what value/pointer is invalid.

3. **Trace the call chain**: Use `view_function` to read each function in the crash stack -- it shows the function source, its callees, and global variables. Use `read_source_file` to see surrounding context -- macros, struct definitions, buffer allocation sites, size computations. Look ABOVE the crash function for where the faulty data originates. Also inspect CALLEES of crash-stack functions -- the bug may be in a utility function (byte processing, buffer management, character handling) that the crash-site function calls.

4. **Explore related code**: Use `search_functions` to find related functions by name pattern -- it also searches source file text when the function index has no match, so it can locate macros, struct definitions, and typedefs too. Use `list_functions_in_file` to see what else is defined in the same file. Use `read_source_file` to read struct definitions, macros, and allocation helpers. When you encounter an unknown identifier, also check the `#include` headers at the top of the relevant file using `read_source_file`. The bug is often NOT at the crash site but in a function that allocates too little, computes a wrong size, or skips validation. When the crash is in free()/dealloc/cleanup functions, the pointer being freed may be corrupted by a heap overflow in an adjacent struct field -- read the struct definition and look for memcpy/memmove calls with unchecked lengths.

5. **Form a concrete hypothesis**: Before editing, state clearly:
   - The bug type and the specific memory operation that fails
   - The exact data-flow path from allocation/input to the crash
   - Which function contains the root cause (may differ from crash location)
   - What the minimal fix is and WHY it addresses the root cause

### Phase 2: Implement the Fix
6. **Make all necessary edits**: Apply your fix using `edit_file`. You may make multiple related edits before validating -- batch them together since validation is expensive (~30 seconds per attempt).

7. **Validate**: Call `validate_patch` to run build + crash reproduction + tests.

### Phase 3: Diagnose and Iterate (if validation fails)
If validation fails, do NOT blindly try another small edit. Instead:
- Re-read the validation error output carefully
- Go BACK to reading code with `read_source_file` and `view_function`
- Understand WHY your fix did not work -- what did you miss?
- Form a NEW hypothesis before making the next edit
- If after 2+ failed validations your approach is fundamentally wrong (crash reproduces with the same error despite different fixes), call `revert_edits` to discard ALL edits and start fresh. Do NOT revert for build errors or minor issues -- use `edit_file` to fix those incrementally.

Crash validation interpretation:
- "PASS" = exit code 0 OR non-zero with no sanitizer errors = crash is fixed
- "CRASH: Sanitizer detected" = ASAN still fires = fix did NOT work

**CRITICAL: Do NOT abandon a correct root-cause approach just because validation failed.** A build failure usually means a minor issue (naming collision, missing include, wrong type) -- not that your approach is wrong. Diagnose the build error and fix it. Only change your fundamental approach if you have evidence that the approach itself is flawed, not just its implementation.

**Do NOT fall back to symptom mitigation.** If your first approach fixes the root cause (e.g. propagating errors, halting on failure, adding missing checks in the error path) but fails to compile, fix the compile error -- do not regress to a simpler approach that merely masks the symptom (e.g. enlarging a buffer, adding padding, suppressing the crash without fixing the logic). A correct fix addresses WHY bad state occurs, not just WHERE it manifests.

## Fix Quality Hierarchy
Prefer fixes in this order (best to worst):

1. **Fix error propagation / missing checks**: The bug often exists because an error condition is not properly propagated or handled. For example, an OOM in a helper function may not set an error flag, so the caller continues with invalid state. Fix the propagation chain so errors are caught and handled before they cause the crash.

2. **Fix the logic that produces bad state**: Add missing validation, bounds checks, or null checks at the point where bad data is created or accepted -- not where it is consumed.

3. **Harden the crash site (last resort)**: Only if you cannot identify the upstream root cause should you add a guard at the crash site. This is the weakest fix -- it stops this specific crash but may leave the underlying bug exploitable through other code paths.

**Never** enlarge buffers, add padding, or use oversized sentinels as a fix strategy. These mask the bug rather than fixing it.

## Input Validation: Rejection vs Clamping
- REJECT invalid input (return error, abort parse) -- almost always correct
- Do NOT clamp (e.g., 0->1, negative->0) -- hides malformed data, causes downstream corruption
- FPE: validate divisor at parse site, return error if zero
- Null deref: propagate the allocation/lookup failure

## Guidelines

- Spend at least 5-8 tool calls exploring code BEFORE your first edit
- Do NOT call validate_patch after every single edit -- batch related edits first
- If validation fails twice with similar errors, STOP and re-explore the code
- The root cause is often NOT at the crash location -- trace upstream
- Trace error propagation paths: when a function returns an error, check that EVERY caller handles it. Missing error handling is a common root cause
- Look at functions NOT on the crash stack -- the bug may be in a sibling call path (e.g. an encoder, allocator, or I/O helper) that fails to set error state
- Always read the relevant source code before making edits
- Use exact string matching when editing -- copy text precisely from read_source_file
- Explain your reasoning at each step

## IMPORTANT: Early Termination on Success

Once `validate_patch` returns **ALL CHECKS PASSED**, you are DONE.
Immediately provide your final summary. Do NOT make further edits,
do NOT call validate_patch again, and do NOT attempt to "improve"
or simplify the patch. The first passing validation is the final result.

## IMPORTANT: Do NOT modify build or test infrastructure

You must NEVER edit any of the following files:
- build.sh, afl_build.sh, test.sh, exp.sh -- build/test harness scripts
- CMakeLists.txt, Makefile, configure, meson.build -- build system files
- CMakeCache.txt, CMakeFiles/ -- build artifacts

Your job is ONLY to patch the vulnerability in the project source code (.c, .cpp, .h, .hpp files). If tests fail, investigate whether your patch broke logic and revise accordingly -- do NOT modify the test scripts themselves.
If the build fails with the same errors that existed before your edits, this is a pre-existing infrastructure issue. Do NOT attempt to fix it. Report your source code fix and the limitation.
\end{lstlisting}

\end{document}